\global\def\draftcontrol{0}

   \def\versionno{ General BCFW: Consistency -- draft   }

\catcode`\@=11

\expandafter\ifx\csname draftcontrol\endcsname\relax\global\def\draftcontrol{0}
\fi

{\count255=\time\divide\count255 by 60
\xdef\hourmin{\number\count255}
\multiply\count255 by-60\advance\count255 by\time
\xdef\hourmin{\hourmin:\ifnum\count255<10 0\fi\the\count255}}
\def\draftdate{\number\month/\number\day/\number\year\ \ \ \hourmin }

\newcommand\makepapertitle{\par
  \begingroup
    \renewcommand\thefootnote{\@fnsymbol\c@footnote}%
    \def\@makefnmark{\rlap{\@textsuperscript{\normalfont\@thefnmark}}}%
    \long\def\@makefntext##1{\parindent 1em\noindent
            \hb@xt@1.8em{%
                \hss\@textsuperscript{\normalfont\@thefnmark}}##1}%
     \newpage
     \global\@topnum\z@   
     \@makepapertitle
     \thispagestyle{empty}\@thanks
  \endgroup
  \setcounter{footnote}{0}%
  \global\let\thanks\relax
  \global\let\makepapertitle\relax
  \global\let\@makepapertitle\relax
  \global\let\@thanks\@empty
  \global\let\@author\@empty
  \global\let\@date\@empty
  \global\let\@title\@empty
  \global\let\title\relax
  \global\let\author\relax
  \global\let\date\relax
  \global\let\and\relax
  \def\version{\let\version\@version\@gobble}
}
\def\@makepapertitle{%
  \newpage
   \ifnum\draftcontrol=1 {}
   \version\versionno
   \vskip 3em%
   \else
   \hfill\hbox to 3cm {\parbox{4cm}{\@pubnum}\hss}%
   \vskip 3em%
   \fi
   \begin{center}%
   \let \footnote \thanks
     {\LARGE {\@title}}%
     \vskip 1.5em%
     {\normalsize
       \lineskip .5em%
       \begin{tabular}[t]{c}%
         \@author
       \end{tabular}\par}%
     \vskip 1.5em%
     {\@bstract}%
     \end{center}%
     \vskip 1.5em
     \@date%
   \par
}

\gdef\@pubnum{}
\def\pubnum#1{%
  \gdef\@pubnum{#1}}

\gdef\@bstract{}
\def\Abstract#1{%
  \gdef\@bstract{%
   \parbox{\textwidth-0pc}{%
   \centerline{\bf Abstract}\penalty1000%
\kern.2cm%
\noindent
\renewcommand\baselinestretch{1.0}%
{#1}}}
}

\def\ps@paper{\let\@mkboth\@gobbletwo%
     \ifnum\draftcontrol=1
    \def\@oddfoot{\hbox to \textwidth{\tiny \versionno \hfil\tiny\draftdate}%
    \hskip -\textwidth \hbox to \textwidth{\hfil\rm\thepage\hfil}}%
     \else\def\@oddfoot{\hbox to \textwidth{\hfil\rm\thepage\hfil}}
     \fi
     \let\@evenfoot\@oddfoot
}

\def\body{\clearpage
          \pagestyle{paper}
    }

\def\@version#1{\ifnum\draftcontrol=1
\typeout{}\typeout{#1}\typeout{}
\vskip3mm\centerline{\hbox{\fbox{\normalsize{\tt DRAFT -- #1 -- }
                   {\draftdate}}}}\vskip3mm
\fi}
\let\version\@version
\long\def\eqlabel#1{\ifnum\draftcontrol=1
                    \tag@false  
                    \tag*{(\theequation) \hbox to -0.2cm{\hspace{0cm}\small{#1}\hss}}
                    \refstepcounter{equation}
                    \edef\@currentlabel{\theequation}
                    \ltx@label{#1}          
                    \else
                    \label{#1}
                    \fi
                    }
\let\st@bibitem\@bibitem
\let\st@lbibitem\@lbibitem
\ifnum\draftcontrol=1
  \def\@bibitem#1{%
    \st@bibitem{#1}\a@@label{#1}\ignorespaces}
  \def\@lbibitem[#1]#2{%
    \st@lbibitem[#1]{#2}\a@@label{#2}\ignorespaces}
  \def\a@@label#1{%
    \gdef\a@lab{\smash{\normalfont\small#1}}
    \ifvmode
      \if@inlabel
        \global\setbox\@labels\hbox{%
          \llap{\a@lab\let\a@lab\relax
                \kern\@totalleftmargin\kern\marginparsep}%
          \box\@labels}%
      \fi
    \fi}
\fi

\documentclass[11pt,A4paper]{article}

\usepackage{amsmath,amssymb,array,calc,epsfig}
\usepackage{cite}
\usepackage[
      colorlinks=true,
      linkcolor=red,  
      urlcolor=blue,    
      filecolor=red,     
      linkcolor=blue,
      citecolor = red,
      pdfstartview=FitV,
      pdftitle={},
        pdfauthor={Paolo Benincasa},
        pdfsubject={},
        pdfkeywords={},
        pdfpagemode=None,
        bookmarksopen=true    
      ]{hyperref}

\usepackage{colortbl}
\usepackage{multirow,bigdelim}
\usepackage{arydshln}     
\usepackage{graphicx}
\usepackage{epstopdf}
\usepackage{ccaption}
\usepackage{slashbox, pict2e}

\usepackage{array}
\usepackage{makecell}
\newcolumntype{x}[1]{>{\centering\arraybackslash}p{#1}}

\usepackage{tikz}

\ifnum\draftcontrol=1
\fi

\renewcommand\baselinestretch{1.25}
\renewcommand\section{\@startsection {section}{1}{\z@}%
                                   {-3.5ex \@plus -1ex \@minus -.2ex}%
                                   {2.3ex \@plus.2ex}%
                                   {\normalfont\large\bfseries}}
\renewcommand\subsection{\@startsection{subsection}{2}{\z@}%
                                   {-3.25ex\@plus -1ex \@minus -.2ex}%
                                   {1.5ex \@plus .2ex}%
                                   {\normalfont\normalsize\bfseries}}
\renewcommand\subsubsection{\@startsection{subsubsection}{3}{\z@}%
                                   {-3.25ex\@plus -1ex \@minus -.2ex}%
                                   {1.5ex \@plus .2ex}%
                                   {\normalfont\normalsize\it}}
\renewcommand\paragraph{\@startsection{paragraph}{4}{\z@}%
                                   {-3.25ex\@plus -1ex \@minus -.2ex}%
                                   {1.5ex \@plus .2ex}%
                                   {\normalfont\normalsize\bf}}

\numberwithin{equation}{section}

\def\revise#1       {\raisebox{-0em}{\rule{3pt}{1em}}%
                     \marginpar{\raisebox{.5em}{\vrule width3pt\
                     \vrule width0pt height 0pt depth0.5em
                     \hbox to 0cm{\hspace{0cm}{%
                     \parbox[t]{4em}{\raggedright\footnotesize{#1}}}\hss}}}}

\def\sqr#1#2{{\vcenter{\vbox{\hrule height.#2pt
 \hbox{\vrule width.#2pt height#1pt \kern#1pt
 \vrule width.#2pt}\hrule height.#2pt}}}}

\def\aa1{\phi}
\def\cc1{\psi}

\catcode`\@=12

\begin{document}


\title{\bf Exploring the S-Matrix of Massless Particles}

\pubnum{%
arXiv:1108.xxxx}
\date{August 2011}

\author{
\scshape Paolo Benincasa${}^{\dagger}$, Eduardo Conde$^{\ddagger}$\\[0.4cm]
\ttfamily Departamento de F{\'i}sica de Part{\'i}culas, Universidade de Santiago de Compostela\\
\ttfamily and\\
\ttfamily Instituto Galego de F{\'i}sica de Altas Enerx{\'i}as (IGFAE)\\
\ttfamily E-15782 Santiago de Compostela, Spain\\[0.2cm]
\small \ttfamily ${}^{\dagger}$paolo.benincasa@usc.es, ${}^{\ddagger}$eduardo@fpaxp1.usc.es
}

\Abstract{We use the recently proposed generalised on-shell representation for scattering amplitudes and a consistency
 test to explore the space of tree-level consistent couplings in four-dimensional Minkowski spacetime. The extension of the 
 constructible notion implied by the generalised on-shell representation, {\it i.e.}  the possibility to reconstruct at tree
 level all the scattering amplitudes from the three-particle ones, together with the imposition of the consistency 
 conditions at four-particle level, allow to rediscover all the known theories and their algebra structure, if any. 
 Interestingly, this analysis seems to leave room for high-spin couplings, provided that at least the requirement of 
 locality is weakened. We do not claim to have found tree-level consistent high-spin theories, but rather that our methods 
 show signatures of them and very likely, with a suitable modification, they can be a good framework to perform a systematic
 search.
}

\makepapertitle

\body

\version\versionno

\section{Introduction}
\label{sec:Intro}

Perturbation theory in particle physics has been mainly understood via Feynman diagrams, which are diagrammatic rules
related to a Lagrangian formulation of the theory.  Such a representation makes manifest properties such as Lorentz 
invariance as well as locality of the interactions. The price one pays is that many other aspects of the theory are hidden: 
in the case of gauge theories, for example, the individual Feynman diagrams break gauge invariance and more generally may 
hide other symmetries of the theory.

A signature of the possible existence of a simple structure for the perturbation theory was already encoded in
formulas based on the Berends-Giele recursion relation 
\cite{Berends:1987me, Mangano:1987xk, Berends:1988zp, Mangano:1990by, Berends:1989hf, 
Kosower:1989xy}, as well as the Parke-Taylor formula for MHV\footnote{As usual, with MHV (Maximally Helicity
Violating) we indicate amplitudes with two negative helicity particles and $n-2$ positive helicity ones.} 
amplitudes of gluons \cite{Parke:1986gb}:
\begin{equation}\eqlabel{ParkeTaylor}
 M_{n}\left(1^{\mbox{\tiny $+$}},\ldots,i^{\mbox{\tiny $-$}},\ldots,j^{\mbox{\tiny $-$}},\ldots,
  n^{\mbox{\tiny $+$}}\right)\:=\:\frac{\langle i,j\rangle^4}{\prod_{k=1}^{n}\langle k,\,k+1\rangle},
 \qquad n+1\,\equiv\,1.
\end{equation}
The Parke-Taylor formula \eqref{ParkeTaylor} has a very simple expression which is not at all manifest in
the Feynman diagrams representation: for a high number $n$ of external gluons, there would be a huge amount
of Feynman diagrams to be summed to obtain the simple answer \eqref{ParkeTaylor}. This is indeed a clear indication
that Feynman diagrams show a large amount of redundancy and hide an intrinsic simplicity of the scattering amplitudes.

Recently, further progress has been made in understanding the perturbative structure of field theories.
Specifically, new representations of the tree-level amplitudes have been found, from the CSW rules
\cite{Cachazo:2004kj} to new sets of on-shell recursion relations (BCFW construction) in Yang-Mills 
\cite{Britto:2004aa, Britto:2005aa}, gravity \cite{Bedford:2005yy, Benincasa:2007qj}, maximally supersymmetric theories
\cite{ArkaniHamed:2008gz} and theories with several type of particles whose highest spin is one or two \cite{Cheung:2008dn},
to the Grassmannian representation for scattering amplitudes in $\mathcal{N}=4$ SYM, which encodes both the tree level and loops
\cite{ArkaniHamed:2009dn, ArkaniHamed:2009sx}.

In the CSW case, the amplitudes are expressed in terms of diagrams containing just MHV-vertices with a light-like
direction suitably chosen \cite{Cachazo:2004kj}. The resulting representation turns out to break Lorentz invariance
at intermediate stages while preserving locality. Unfortunately, such a construction is well-defined just in the case
of Yang-Mills theory\footnote{An attempt to prove an MHV-expansion for gravity amplitudes was made in 
\cite{BjerrumBohr:2005jr} using the BCFW method. However, already in the case of NMHV amplitudes, such a construction turns 
out to be valid just for less than $12$ external states, as proven numerically in \cite{Bianchi:2008pu} and analytically in
\cite{Benincasa:2007qj}.}.

The on-shell recursion relations, instead, can be straightforwardly obtained by using the power of complex analysis, 
as suggested in \cite{Britto:2005aa} for Yang-Mills tree-level amplitudes. The idea of \cite{Britto:2005aa} is to introduce 
a one-parameter deformation of the complexified momentum space, which generates a one-parameter family of 
amplitudes, and to reconstruct the physical amplitude from its singularity structure. The procedure
is general and can be used to look for recursive structures in any theory. Furthermore, the recursive structure itself
makes Lorentz-invariance and gauge-invariance manifest, at the price of breaking locality in the individual on-shell 
diagrams.

Another issue which points to our incomplete understanding of the field theory structure is the question of whether it is possible to
formulate high-spin interactions in flat space-time. The free theory is fairly well understood (see the reviews
\cite{Sorokin:2004ie, Bouatta:2004kk, Francia:2006hp, Fotopoulos:2008ka} and references therein). In particular, the impossibility of massless particles with spin $\,\ge\,3$ to mediate long-range 
forces was established in \cite{Weinberg:1964ew}, by looking at soft emissions in the S-matrix containing scalars and a single high-spin particle (for a 
generalisation of this analysis to include external arbitrary spin particles, see \cite{Taronna:2011kt}). One of the 
implicit assumptions in such an argument is the local nature of the couplings, and therefore the soft limit analysis
in the context of local interacting theories puts strong constraints on the high-spin interactions, not forbidding them in 
general but rather ruling out the possibility that high spins may produce macroscopic effects. 

One of the biggest problems in formulating a consistent interacting theory has been the construction of couplings which 
preserve the full high-spin gauge invariance. However, several attempts have been made to construct consistent interactions
\cite{Berends:1984wp, Fradkin:1987ks, Fradkin:1986qy, Deser:1990bk, Henneaux:1997bm, Bekaert:2006us, Fotopoulos:2007nm, 
      Fotopoulos:2007yq, Boulanger:2008tg, Bekaert:2009ud, Francia:2010qp, Bekaert:2010hp, Taronna:2011kt}.
An indication that the requirement of locality must be possibly dropped already comes from a geometric formulation of the free 
theory, which returns non-local equations for high-spin particles \cite{Francia:2002aa, Francia:2002pt}, as well as from 
further studies on the interactions ({\it e.g.} \cite{Francia:2010qp, Taronna:2011kt}).

In the pursuit of a deeper understanding of the perturbative structure of interacting theories - starting with 
the tree-level approximation for theories of massless particles -, ideally one would like to formulate a general
S-Matrix theory starting from a minimal amount of assumptions. A good starting point seems to be the BCFW construction.
The generality of the procedure as well as the possibility of imposing a well-defined set of consistency conditions on the 
S-Matrix \cite{Benincasa:2007xk}, seem to suggest four fundamental assumptions: analyticity (which guarantees that the 
singularity structure of the S-Matrix is characterised by just poles and branch cuts), Poincar{\'e} invariance\footnote{
Generically speaking, this assumption can be substituted by the invariance under the Super-Poincar{\'e} group in case of
supersymmetric theories, as in \cite{ArkaniHamed:2008gz} for maximally supersymmetric theories, or, it can be further 
generalised to the isometry group of the space-time whose irreducible representations define the asymptotic states.}(which
defines the asymptotic states through its irreducible representations), the existence of one-particle states (which 
allows to define operators which act on the amplitudes as they act on the individual particles) and the locality of
the whole S-Matrix (which guarantees that the singularities come just from propagators). We will further comment on the possibility of dropping this
last requirement.

The analysis of the singularity structure of the amplitudes to reconstruct them is not in itself a new approach,  but it can be traced back to the S-matrix program of the 60's \cite{Olive:1964aa, Chew:1966aa, Eden:1966ab}. However, there 
are striking differences between the S-matrix program analysis and the method proposed in \cite{Britto:2005aa}. First,
in the 60's the focus was on massive particles with spin less than one. Second, the analysis of the 
singularity structure was treated as a multi-variable problem in terms of Lorentz invariant quantities: in this way the 
number of variables dramatically increases with the growth of the number of external states. These two features created
a big obstruction to the complete realisation of the S-matrix program. The introduction of a one-parameter
deformation allows to consider the amplitude just as a function of such a single parameter, and therefore to analyse the 
singularities in a much simpler way.

Restricting ourselves to the analysis of the pole structure is equivalent to considering the tree-level approximation for 
the amplitudes. Thus, reconstructing  scattering amplitudes from their pole structure means relating them to the residues of
such poles. This is indeed possible if the deformed amplitudes vanish as momenta are taken to infinity along some complex
direction, as it was shown to be the case in \cite{Britto:2005aa} for an arbitrary number of external gluons, and in 
\cite{Benincasa:2007qj} for an arbitrary number of external gravitons. The residues are just products of
two on-shell scattering amplitudes with fewer external states. An on-shell amplitude may therefore be related to on-shell 
amplitudes with fewer external particles, providing a recursion relation.

If one iterates such a relation, one can  discover that, independently of the vertex structure of the Lagrangian formulation of the theory, any scattering amplitude is determined by three-particle amplitudes. It is well known that on-shell 
three-particle amplitudes of massless particles vanish in Minkowski space. However, the procedure illustrated above is 
defined in the complexified Minkowski space, where they do not vanish \cite{Witten:2003aa}. Therefore, one can state that 
the tree level of theories with such a recursive structure is fully determined by the knowledge of the three-particle 
amplitudes. For this reason theories having this feature, such as pure Yang-Mills and General Relativity, have been called
 {\it fully constructible} \cite{Benincasa:2007xk}. Maximally supersymmetric theories in four dimensions, {\it i.e.} 
$\mathcal{N}=4$ SYM and $\mathcal{N}=8$ Supergravity, belong to this class of theories as well \cite{ArkaniHamed:2008gz}.
For theories whose smallest vertex interaction is a $k$-point vertex ($k\,>\,3$), the recursive relations imply that the
$n$-particle amplitude is determined by the $k$-particle amplitudes. However, this class of theories can be treated on
the same footing of the previous ones by defining effective three-particle amplitudes through the introduction of a 
massive particle, which needs to be integrated out after the computation, as it was done for $\lambda\phi^4$
in \cite{Benincasa:2007xk}.

Both from a Lagrangian formulation \cite{ArkaniHamed:2008yf} and from purely S-Matrix arguments \cite{Benincasa:2011kn},
it has been possible to prove that the behaviour of an amplitude as the momenta are taken to infinity along
some complex direction identified by the momenta of two particles, does not depend on the number of external states.
This implies that if an amplitude vanishes at infinity in this limit for some fixed number $n$ of external states, it will 
vanish for any $n$ and, therefore, the corresponding theory is tree-level constructible.

Similarly, if such a condition is not satisfied for some given $n$, it will not be satisfied for any $n$. Physically,
the independence of the number of external states in the (complex)-UV behaviour is understandable if one 
thinks about this limit as a hard particle going through a soft background \cite{ArkaniHamed:2008yf, Cheung:2008dn}.

\begin{figure}[htbp]
 \centering 
  \scalebox{.30}{\includegraphics{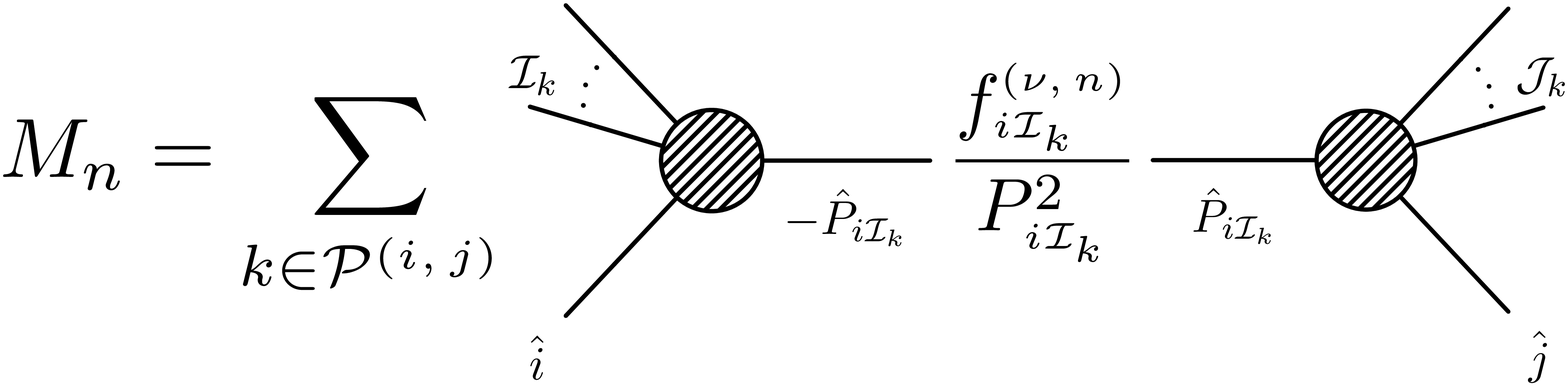}}
  \caption{Generalised on-shell recursion relation. This new recursion relation shows the same structure of the usual
           BCFW one, with an additional factor which depends on a sub-set of the zeroes of the amplitude. The notation will become clear later.}\label{fig:wBCFW}
\end{figure}

The case in which the previous constructibility condition is not satisfied was studied in very specific cases in
\cite{Feng:2009ei, Feng:2010ku}. However, the notion of constructibility has been generalised to include any consistent
theory in \cite{Benincasa:2011kn}. More precisely, in \cite{Benincasa:2011kn} it was shown that the scattering
amplitudes of a generic theory have the BCFW-structure, {\it i.e.}: they can be expressed in terms of products of 
amplitudes with a smaller number of external states, weighted by factors which are one in the case where the amplitude
vanishes as the momenta of the deformed particles are taken to infinity along a complex direction,
while, in the other case, they depend on the channel momentum evaluated at the location of a subset of the zeroes of the 
amplitudes (Fig.: \ref{fig:wBCFW}).

In principle, these ``weights'' may be fixed by looking at the factorisation properties of the full amplitude. Doing that 
suggests a connection between the ``weights'' of a given amplitude and the ones for an amplitude with less external states. 
Even if the conditions on the ``weights'' can be solved in a number of examples, this connection has not been made precise
yet.

In this paper we show how these ``weights'' can be fixed for generic four-particle amplitudes. Furthermore, equipped
with this prescription as well as with the four-particle test proposed in \cite{Benincasa:2007xk}, it is possible
to explore the space of theories with a consistent S-Matrix, at least at tree level. The idea of the four-particle test
is to use two different BCFW-deformations to compute a certain four-particle amplitude and then impose the equality
between the two results \cite{Benincasa:2007xk}. If the theory is consistent, the equality has to hold given that a physical
quantity as the scattering amplitude must not depend on the particular BCFW-deformation chosen to compute it. In \cite{Benincasa:2007xk}, 
such a test could be applied just to theories having amplitudes for which it was possible to choose two BCFW-deformations 
such that they were both vanishing in the (complex)-UV limit. Even if this was allowing to scan only a subset of all possible 
theories, this constraint was returning very interesting results, {\it i.e.}: the Jacobi Identity in Yang-Mills was not
imposed a priori but it was arising as a consequence of the consistency condition, as well as the supersymmetry in 
$\mathcal{N}=1$ supergravity.

In the present case, the generalised on-shell representation \cite{Benincasa:2011kn} allows us to complete the scan of 
four-dimensional interacting theories which admit a definition of asymptotic states. In particular, it is now possible to 
detect theories such as $\lambda\phi^3$, Einstein-Maxwell, Scalar-Gravity, Scalar-QED, etc. Furthermore, the 
tree-level consistency conditions seem to allow for the existence of a self-interacting spin-$2$ theory with $6$-derivative
three-particle interactions and particular theories involving particles with spin higher than $2$. As mentioned earlier,
the no-go theorem about the non-existence of high-spin theories in Minkowski space is related to the impossibility 
of writing down interaction terms preserving the high-spin gauge symmetry of the theory, and to the fact that they
cannot mediate long-range forces. These no-go theorems can be avoided if the basic assumption of locality is dropped.
Now, the BCFW construction breaks locality at intermediate stages: the individual on-shell diagrams show spurious
singularities which disappear when they get summed. In this sense, the BCFW construction might be a good framework to
look for high-spin theories, if we learn how to drop the locality requirement for the full S-matrix.

The paper is organised as follows. In Section \ref{sec:BCFWrev} we review the very basic features of the S-Matrix, the BCFW 
construction and its generalisation to theories whose amplitudes do not vanish as the momenta are taken to infinity along
some complex direction (complex-UV limit). We further comment on the notion of constructibility, proposing and discussing 
the equivalence between the complex-UV behaviour and the soft-behaviour of three-particle amplitudes. In Section 
\ref{sec:4pt2} we focus on the four-particle amplitudes. We rewrite the conditions on the zeroes in a very compact way and 
use the generalised on-shell recursion relations to scan the space of non-trivial interactive theories in four dimensions. 
We mainly focus on theories whose coupling constants have all the same dimensionality\footnote{In a Lagrangian description 
this is equivalent to considering theories whose three-particle vertices are characterised by the same number of derivatives.}.
We classify the theories according to the dimensionality of the three-particle coupling constant, and for each of these classes
we discuss the implications of the consistency requirement. In Section \ref{sec:Loc} we discuss our result from the S-matrix 
consistency condition in relation to the scattering of particles of spin higher than two and we comment on the requirement
of locality. Finally, Section \ref{sec:Concl} contains the conclusion and further comments.

\section{The S-Matrix and the BCFW construction}
\label{sec:BCFWrev}

In this section we review the basic properties of the S-Matrix and the BCFW construction. In particular,
we want to argue that the BCFW representation allows for a reformulation of the S-matrix theory in terms of just four 
fundamental hypothesis:
\begin{enumerate}
 \item Poincar{\'e} invariance.
 \item Existence of one-particle states.
 \item Locality of the full S-Matrix.
 \item Analyticity.
\end{enumerate}

\subsection{S-Matrix}
\label{subsec:SM}

The S-Matrix elements are the transition amplitudes for $m$ asymptotic initial states scattering into $n-m$ final asymptotic 
states. The scattering amplitudes are obtained from these by stripping the identity matrix from the S-matrix, that accounts
for the possibility of the interaction to be trivial.
The asymptotic states are defined as the irreducible representations of the space-time isometry group. Given 
that we are considering scattering in four-dimensional\footnote{Here we focus on the S-Matrix in four dimensions
and we make use of the four-dimensional helicity-spinor formalism. However, a generalisation of such techniques to
different dimensions has been proposed in \cite{Cheung:2009dc, Boels:2009bv, CaronHuot:2010rj, Gang:2010gy}.} 
Minkowski space, such an isometry group is the Poincar{\'e} group $\mathbb{P}_{4}\,=\,\mathcal{T}^{4}\ltimes SO(3,1)$. 

In the case of massless representations of the Poincar{\'e} group, the physical information is encoded into the 
null momentum $p^{\mbox{\tiny $(i)$}}$ and the polarisation tensors $\varepsilon_{\mu_1\ldots\mu_s}^{(i)}$  of the particles, or 
equivalently in the pair of spinors $(\lambda^{\mbox{\tiny (i)}}_{a},\,\tilde{\lambda}^{\mbox{\tiny (i)}}_{\dot{a}})$ and 
the helicity $h_{i}\,=\,\pm s$, $s$ being the spin of the particle. This equivalence holds
because of the isomorphism $SO(3,1)\,\cong\,SL(2,\mathbb{C})$, which is implemented by the Pauli matrices
$\sigma^{\mu}_{a\dot{a}}\,=\,\left(\mathbb{I}_{a\dot{a}},\,\overrightarrow{\sigma}_{a\dot{a}}\right)$:
\begin{equation}\eqlabel{iso}
 p_{\mu}\quad\longrightarrow\quad p_{a\dot{a}}\:=\:\sigma^{\mu}_{a\dot{a}}p_{\mu}
  \:=\:\lambda_{a}\tilde{\lambda}_{\dot{a}},
\end{equation}
where the last equality is possible whenever $p_{\mu}$ is null, and $\lambda_{a}$ and $\tilde{\lambda}_{\dot{a}}$ transform respectively in the $(1/2,\,0)$ and 
$(0,\,1/2)$ representation of $SL(2,\mathbb{C})$. It is possible to define two inner product for spinors,
one for each representation of $SL(2,\mathbb{C})$ under which they can transform:
\begin{equation}\eqlabel{innprod}
 \langle\lambda,\,\lambda'\rangle\:\equiv\:\epsilon^{ab}\lambda_{a}\lambda'_{b},\qquad
 [\tilde{\lambda},\,\tilde{\lambda}']\:\equiv\:\epsilon^{\dot{a}\dot{b}}
  \tilde{\lambda}_{\dot{a}}\tilde{\lambda}'_{\dot{b}},
\end{equation}
with $\epsilon_{12}\,=\,1\,=\,\epsilon_{\dot{1}\dot{2}}$, $\epsilon^{12}\,=\,-1\,=\,\epsilon{\dot{1}\dot{2}}$,
and $\epsilon^{ac}\epsilon_{cb}\,=\,\delta^{a}_{\phantom{a}b}$. Notice that the inner products \eqref{innprod}
are Lorentz invariant. 

Assuming the existence of one-particle states and that the Poincar{\'e} group acts on scattering amplitudes
as it acts on individual states implies the following action of the helicity operator
\begin{equation}\eqlabel{helop}
 \left(\lambda^{\mbox{\tiny $\left(i\right)$}}_{a}
  \frac{\partial}{\partial\lambda^{\mbox{\tiny $\left(i\right)$}}_{a}}-
 \tilde{\lambda}^{\mbox{\tiny $\left(i\right)$}}_{\dot{a}}
  \frac{\partial}{\partial\tilde{\lambda}^{\mbox{\tiny $\left(i\right)$}}_{\dot{a}}}
 \right)M_{n}\left(1^{\mbox{\tiny $h_{1}$}},\ldots,n^{\mbox{\tiny $h_{n}$}}\right)\:=\:-2h_{i}\,
  M_{n}\left(1^{\mbox{\tiny $h_{1}$}},\ldots,n^{\mbox{\tiny $h_{n}$}}\right).
\end{equation}

At tree level, the scattering amplitudes are rational functions of Lorentz
invariants quantities, since the singularities can only be poles. Requiring the locality of the theory implies that the poles must come just from 
internal propagators, {\it i.e.} they correspond to virtual massless particles going on-shell.

\subsection{BCFW construction}
\label{subsec:BCFW}

In the complexified momentum space ($p^{\mbox{\tiny $(i)$}}\,\in\,\mathbb{C}^4$), it is possible to introduce 
a one-complex-parameter deformation such that the deformed momenta satisfy both the on-shell condition and
the momentum conservation, defining a one-parameter family of amplitudes. Such a
deformation is not defined univocally. The simplest example may be defined by deforming the momenta of two 
particles only \cite{Britto:2005aa}, leaving the others unchanged
\begin{equation}\eqlabel{BCFWdef}
 p^{\mbox{\tiny $(i)$}}\:\longrightarrow\:p^{\mbox{\tiny $(i)$}}(z)\,=\,p^{\mbox{\tiny $(i)$}}-zq,
 \qquad
 p^{\mbox{\tiny $(j)$}}\:\longrightarrow\:p^{\mbox{\tiny $(j)$}}(z)\,=\,p^{\mbox{\tiny $(j)$}}+zq,
 \qquad
 p^{\mbox{\tiny $(k)$}}\:\longrightarrow\:p^{\mbox{\tiny $(k)$}}(z)\,=\,p^{\mbox{\tiny $(k)$}},
\end{equation}
where $k$ can label all the particles except for $i$ and $j$. It is straightforward to see that
the deformation \eqref{BCFWdef} satisfies the momentum conservation condition. The requirement that the
deformed momenta are still on-shell fixes $q$ to be a complex vector in $\mathbb{C}^4$
\begin{equation}\eqlabel{q}
 q^2\,=\,0,\qquad p^{\mbox{\tiny $\left(i\right)$}}\cdot q\,=\,0\,=\,p^{\mbox{\tiny $\left(j\right)$}}\cdot q.
\end{equation}
A deformation of this type defines a one-parameter family of physical amplitudes $M_{n}^{\mbox{\tiny $(i,j)$}}(z)$. The 
multi-variable problem is therefore mapped into a single variable problem, {\it i.e.} it is now possible to 
analyse the singularity structure of the amplitude as a function of $z$. The propagators correspond to simple 
poles of the form:
\begin{equation}\eqlabel{poleZ}
 \frac{1}{\left[P_{k}(z)\right]^2}\:=\:\frac{1}{P_{k}^2-2z\left(P_{k}\cdot q\right)}
 \quad\Rightarrow\quad
 z\rightarrow z_{k}=\frac{P_{k}^2}{2\left(P_{k}\cdot q\right)},
\end{equation}
where $P_{k}=\sum_{l\in\mathcal{S}_k}p^{\mbox{\tiny $(l)$}}$, $\mathcal{S}_k$ is the set of external states in
the $k$-channel, $i\in\mathcal{S}_k$ and $z_k$ is the location of the pole. One can now imagine to relate
the physical amplitude (which can be obtained from $M_{n}^{\mbox{\tiny $(i,j)$}}(z)$ by setting $z=0$) to the residues of 
the poles in \eqref{poleZ}. Let $\mathcal{R}$ be the Riemann sphere obtained as the union of the complex plane with the 
point at infinity $\mathcal{R}\:=\:\mathbb{C}\cup\left\{\infty\right\}$. Then

\begin{equation}\eqlabel{AmplInt}
 0\:=\:\frac{1}{2\pi i}\oint_{\mathcal{R}}\frac{dz}{z}\:M_{\mbox{\tiny $n$}}^{\mbox{\tiny $(i,\,j)$}}\left(z\right)\:=\:
  M_{\mbox{\tiny $n$}}^{\mbox{\tiny $(i,j)$}}\left(0\right)+
  \sum_{k\in\mathcal{P}^{\mbox{\tiny $(i,j)$}}}\frac{c_{k}^{\mbox{\tiny $(i,j)$}}}{z_{\mbox{\tiny $k$}}}-
   \mathcal{C_{\mbox{\tiny $n$}}^{\mbox{\tiny $(i,\,j)$}}},
\end{equation}
where $M_{\mbox{\tiny $n$}}^{\mbox{\tiny $(i,j)$}}\left(0\right)\,\equiv\,M_n$. Some comments are in order. 

First, in the integral appearing in \eqref{AmplInt} we used the notation 
$M_{\mbox{\tiny $n$}}^{\mbox{\tiny $(i,\,j)$}}(z)$ to specify that the one-parameter family of amplitudes has been obtained 
by deforming the momenta of the particles labelled by $i$ and $j$. Different deformations produce different families of 
amplitudes, all of them containing the physical one which can be obtained by setting the parameter $z$ to zero. The families
of amplitudes that can be defined via different momenta deformations differ in the location of the poles at finite point 
$z_{k}$ (we denote by $\mathcal{P}^{\mbox{\tiny $(i,j)$}}$ the set of poles $\left\{z_{k}\right\}$ created by the deformation
$\left(i,j\right)$). The term $c_{k}^{\mbox{\tiny $(i,j)$}}$ is just the residue of the pole $z_{i}$ for 
$M_{\mbox{\tiny $n$}}$. Finally, $\mathcal{C_{\mbox{\tiny $n$}}^{\mbox{\tiny $(i,\,j)$}}}$ is the contribution from the
point at infinity.

The equation \eqref{AmplInt} therefore relates the physical amplitude $M_{\mbox{\tiny $n$}}$ 
to the residues of the poles $\left\{z_{k}\right\}$ and to $\mathcal{C_{\mbox{\tiny $n$}}^{\mbox{\tiny $(i,\,j)$}}}$. The 
interpretation of the residues $c_{k}^{\mbox{\tiny $(i,j)$}}$ has been provided in \cite{Britto:2005aa} and it is just the 
product of two on-shell amplitudes with fewer external states. In order to understand this, let us consider a specific pole 
$z_{k}$, which appears through an internal propagator as in \eqref{poleZ}. This means that we are focusing on a specific
channel: as $z\,\rightarrow\,z_{k}$ the momenta $P_{k}(z)$ in \eqref{poleZ} goes on-shell, and this channel
dominates on the others. The $n$-point amplitude factorises into the product of two on-shell amplitudes:
\begin{equation}\eqlabel{ci}
 M_{n}^{\mbox{\tiny $(i,j)$}}\left(z\right)\:\overset{\mbox{\tiny $z\,\rightarrow\,z_{k}$}}{\sim}\:
   M_{\mbox{\tiny L}}^{\mbox{\tiny $(i,j)$}}\left(z_{\mbox{\tiny $k$}}\right)
   \frac{1}{2\left(P_{\mbox{\tiny $k$}}\cdot q\right)\left(z_{\mbox{\tiny $k$}}-z\right)}
   M_{\mbox{\tiny R}}^{\mbox{\tiny $(i,j)$}}\left(z_{\mbox{\tiny $k$}}\right).
\end{equation}
The corresponding residue in the sum in \eqref{AmplInt} is simply given by
\begin{equation}\eqlabel{cizi}
 -\frac{c_{k}^{\mbox{\tiny $(i,j)$}}}{z_{k}}\:=\:
  M_{\mbox{\tiny L}}^{\mbox{\tiny $(i,j)$}}\left(z_{\mbox{\tiny $k$}}\right)
   \frac{1}{2\left(P_{\mbox{\tiny $k$}}\cdot q\right)z_{\mbox{\tiny $k$}}}
   M_{\mbox{\tiny R}}^{\mbox{\tiny $(i,j)$}}\left(z_{\mbox{\tiny $k$}}\right)\:=\:
 M_{\mbox{\tiny L}}^{\mbox{\tiny $(i,j)$}}\left(z_{\mbox{\tiny $k$}}\right)
   \frac{1}{P_{\mbox{\tiny $k$}}^2}
   M_{\mbox{\tiny R}}^{\mbox{\tiny $(i,j)$}}\left(z_{\mbox{\tiny $k$}}\right).
\end{equation}
The knowledge of the residues \eqref{cizi} of the poles at finite points is enough to determine the amplitude if and only if
\begin{equation}\eqlabel{constrth}
 \lim_{z\,\rightarrow\,\infty}M_{n}^{\mbox{\tiny $(i,j)$}}(z)\:=\:0,
\end{equation}
which implies that the contribution from infinity $\mathcal{C}_{n}^{\mbox{\tiny $(i,j)$}}$ vanishes and therefore
the physical amplitude is just given by 
\begin{equation}\eqlabel{BCFW-RR}
 M_{\mbox{\tiny $n$}}\,\equiv\,M_{\mbox{\tiny $n$}}^{\mbox{\tiny $(i,j)$}}\left(0\right)\:=\:
   \sum_{k\in\mathcal{P}^{\mbox{\tiny $(i,j)$}}}
   M_{\mbox{\tiny L}}^{\mbox{\tiny $(i,j)$}}\left(z_{\mbox{\tiny $k$}}\right)
   \frac{1}{P_{\mbox{\tiny $k$}}^2}
   M_{\mbox{\tiny R}}^{\mbox{\tiny $(i,j)$}}\left(z_{\mbox{\tiny $k$}}\right).
\end{equation}
The relation \eqref{BCFW-RR} is known as the BCFW recursive relation \cite{Britto:2004aa, Britto:2005aa} and the 
theories which satisfy the condition \eqref{constrth} and therefore admit a BCFW representation 
\eqref{BCFW-RR} are called {\it constructible} \cite{Benincasa:2007xk}. Such a condition is satisfied by
gluons \cite{Britto:2005aa}, gravitons \cite{Benincasa:2007qj}, and maximally supersymmetric theories
\cite{ArkaniHamed:2008gz}.

If the condition \eqref{constrth} is not satisfied, the term $\mathcal{C}_{n}^{\mbox{\tiny $(i,j)$}}$ is needed and the 
knowledge of the poles is no longer enough to determine the scattering amplitude. This problem was discussed in very
particular cases in \cite{Feng:2009ei, Feng:2010ku}, while in \cite{Benincasa:2011kn} a general prescription was
provided which takes into account a subset of the zeroes of the amplitude. We review it in the next subsection.

\subsection{Generalised On-Shell Recursion Relations}\label{subsec:GenRR}

To begin with, let us consider the one-parameter family of amplitudes $M_{n}^{\mbox{\tiny $(i,j)$}}$ generated by
the deformation \eqref{BCFWdef}
\begin{equation}\eqlabel{Mzij}
 M_{n}^{\mbox{\tiny $(i,j)$}}(z)\:=\:\sum_{k\in\mathcal{P}^{\mbox{\tiny $(i,j)$}}}
  \frac{M_{\mbox{\tiny L}}^{\mbox{\tiny $(i,j)$}}(z_k)M_{\mbox{\tiny R}}^{\mbox{\tiny $(i,j)$}}(z_k)}{[P_k(z)]^2}+
 \mathcal{C}_{n}^{\mbox{\tiny $(i,j)$}}(z),
\end{equation}
where the explicit expression for the residues \eqref{cizi} of the poles at finite location was used. The first feature
to notice is that all the poles at finite location are contained in the first term of \eqref{Mzij}, which implies that
$\mathcal{C}_{n}^{\mbox{\tiny $(i,j)$}}(z)$ is just a polynomial of order $\nu$, $\nu$ representing the power with which
\eqref{Mzij} diverges as $z$ is taken to infinity. A further important feature is that, independently of the actual value
of $\nu$ (which is theory- and deformation-dependent), the only contribution from 
$\mathcal{C}_{n}^{\mbox{\tiny $(i,j)$}}(z)$ to the physical amplitude comes from its term of $0$-th order. This is easy to 
understand if one recalls that the physical amplitude is obtained from \eqref{Mzij} for $z=0$: at such a value for $z$, 
$\mathcal{C}_{n}^{\mbox{\tiny $(i,j)$}}(z)$ is reduced to just its $0$-th order term.

Let us now consider a subset $\{z_0^{\mbox{\tiny $(s)$}}\}$ of $n_z$ zeroes of \eqref{Mzij} and contours
$\gamma_0^{\mbox{\tiny $(s)$}}$ containing just the zero $z_0^{\mbox{\tiny $(s)$}}$. Then
\begin{equation}\eqlabel{zeroes}
 \begin{split}
  0\:=\:\frac{1}{2\pi i}\oint_{\mbox{\tiny $\gamma_0^{(s)}$}}dz\:
  &\frac{M_{n}^{\mbox{\tiny $(i,j)$}}(z)}{\left(z-z_0^{\mbox{\tiny $(s)$}}\right)^{r}}\:=\:
  \left(-1\right)^{r-1}\sum_{k=1}^{N_{\mbox{\tiny $P$}}^{\mbox{\tiny fin}}}
  \frac{M_{\mbox{\tiny L}}^{\mbox{\tiny $(i,j)$}}(z_k)M_{\mbox{\tiny R}}^{\mbox{\tiny $(i,j)$}}(z_k)}{
  \left(-2P_k\cdot q\right)\left(z_0^{\mbox{\tiny $(s)$}}-z_k\right)^{r}}+
  \delta_{\mbox{\tiny $r,1$}}\mathcal{C}_{\mbox{\tiny $n$}}^{\mbox{\tiny $(i,j)$}}+\\
  &\hspace{.2cm}+
  \sum_{l=1}^{\nu}\frac{l!}{\left(l-r+1\right)!\left(r-1\right)!}a_{l}^{\mbox{\tiny $(i,j)$}}z_{\mbox{\tiny $0$}}^{l-r+1},
  \quad \mbox{with }
  \left\{
   \begin{array}{l}
    r\,=\,1,\ldots,m^{\mbox{\tiny $(s)$}}\\
    s\,=\,1,\ldots,n_z
   \end{array}
  \right. ,
 \end{split}
\end{equation}
where $N_{\mbox{\tiny $P$}}^{\mbox{\tiny fin}}$ is the number of poles at finite location and $m^{\mbox{\tiny $(s)$}}$ is 
the multiplicity of the zero $z_0^{\mbox{\tiny $(s)$}}$. If $n_z\,=\,\nu+1$, the system of algebraic equations 
\eqref{zeroes} would fix univocally the amplitudes. Solving such a system of equations shows a connection between
$\mathcal{C}_n^{\mbox{\tiny $(i,j)$}}$ and a sum of products of on-shell scattering amplitudes with fewer external states.
The expression in itself is not particularly illuminating, so we will not write it here, rather see \cite{Benincasa:2011kn}.
What is important is that such an expression, once reinserted in \eqref{AmplInt}, allows us to rewrite the scattering amplitudes
in such a way that the overall structure of the BCFW construction is still preserved (Fig.: \ref{fig:wBCFW})
\begin{equation}\eqlabel{GenRR2}
 M_n\:=\:\sum_{k\in\mathcal{P}^{\mbox{\tiny $(i,j)$}}}
  M_{\mbox{\tiny L}}^{\mbox{\tiny $(i,j)$}}(z_k)\frac{f^{\mbox{\tiny $(\nu,n)$}}_k}{P_k^2}
  M_{\mbox{\tiny R}}^{\mbox{\tiny $(i,j)$}}(z_k),
\end{equation}
with the factors $f^{\mbox{\tiny $(\nu,n)$}}_k$ being
\begin{equation}\eqlabel{fweight}
 f_{\mbox{\tiny $k$}}^{\mbox{\tiny $(\nu,n)$}}\:=\:
 \left\{
  \begin{array}{ll}
   1,& \nu\,<\,0,\\
   \phantom{\ldots}\\
   \prod\limits_{l=1}^{\nu+1}\left(1-\frac{P_{\mbox{\tiny $k$}}^2}{P_{\mbox{\tiny $k$}}^2
    \left(z_0^{\mbox{\tiny $(l)$}}\right)}\right),& \nu\,\ge\,0.
  \end{array}
 \right.
\end{equation}
When we say that the BCFW construction is preserved, we mean that the amplitudes still turn out to be
expressed in terms of products of on-shell amplitudes with fewer external states and propagators, but now weighted by
a simple factor which depends on a subset of zeroes of the amplitude.

The recursion relation \eqref{GenRR2} is a valid mathematical expression for the amplitude $M_n$. This means that it has to
factorise properly when collinear/multi-particle limits are taken. Imposing such a requirement, the ``weights'' 
$ f_{\mbox{\tiny $k$}}^{\mbox{\tiny $(\nu,n)$}}$ turn out to satisfy the following conditions:
\begin{equation}\eqlabel{zeroes-sum}
 \begin{split}
  &\hspace{2.5cm}P_{ik}^2(z_0^{\mbox{\tiny $(l)$}})\:=\:\langle i,k\rangle\alpha_{ik}^{\mbox{\tiny $(l)$}}[i,j],\qquad
   P_{jk}^2(z_0^{\mbox{\tiny $(l)$}})\:=\:\langle i,j\rangle\alpha_{jk}^{\mbox{\tiny $(l)$}}[j,k],\\
  &\hspace{2.5cm}\lim_{P_{\mathcal{K}}^2\rightarrow0}f_{\mbox{\tiny $i\mathcal{I}_k$}}^{\mbox{\tiny $(\nu,\,n)$}}\:=\:
    f_{\mbox{\tiny $i\mathcal{I}_k$}}^{\mbox{\tiny $(\nu,\,n-s+1)$}},\qquad
   \lim_{P_{i\mathcal{I}_k}^2\rightarrow0}f_{i\mathcal{I}_k}^{\mbox{\tiny $(\nu,n)$}}\,=\,1,\\
  &\hspace{2.5cm}\lim_{[k_1,k_2]\rightarrow0}f_{\mbox{\tiny $i\bar{k}$}}^{\mbox{\tiny $(\nu,n)$}}\:=\:
    f_{\mbox{\tiny $i(k_1 k_2)$}}^{\mbox{\tiny $(\nu,n-1)$}},\qquad
   \lim_{\langle k_1,k_2\rangle\rightarrow0}f_{\mbox{\tiny $j\bar{k}$}}^{\mbox{\tiny $(\nu,n)$}}\:=\:
    f_{\mbox{\tiny $j(k_1 k_2)$}}^{\mbox{\tiny $(\nu,n-1)$}},\\
  &\lim_{[i,j]\rightarrow0}\sum_{k}(-1)^{2(h_i+h_j+h_k)+\delta+\nu+1}
   \left[
    \left(\frac{\langle i,k\rangle}{\langle i,j\rangle}\right)^{\delta-1}
    \left(\frac{[i,j]}{[i,k]}\right)^{2h_i+\delta-\nu}
    \frac{\mathcal{H}_{n-1}^{(k)}}{\prod_{l=1}^{\nu+1}\alpha_{ik}^{\mbox{\tiny $(l)$}}}\right]\:=\:1,\\
  &\lim_{\langle i,j\rangle\rightarrow0}\sum_k(-1)^{2(h_i+h_k)+\delta+\nu+1}
    \left[
    \left(\frac{[j,k]}{[i,j]}\right)^{\delta-1}
    \left(\frac{\langle i,j\rangle}{\langle j,k\rangle}\right)^{\delta-2h_j-\nu}
    \frac{\tilde{\mathcal{H}}_{n-1}^{(k)}}{\prod_{l=1}^{\nu+1}\alpha_{jk}^{\mbox{\tiny $(l)$}}}
   \right]\:=\:1,
 \end{split}
\end{equation}
where the notation has been detailed in \cite{Benincasa:2011kn}. For the present purposes we only need to know that $\mathcal{H}$ is a dimensionless helicity factor and $\delta$ is the number of derivatives of the three-particle
interactions, and that the BCFW-deformation \eqref{BCFWdef} has been implemented by deforming the anti-holomorphic spinor
for particle-$i$ and the holomorphic one for particle-$j$
\begin{equation}\eqlabel{BCFWdef2}
 \tilde{\lambda}^{\mbox{\tiny $(i)$}}(z)\:=\:\tilde{\lambda}^{\mbox{\tiny $(i)$}}-z\tilde{\lambda}^{\mbox{\tiny $(j)$}},
 \quad
 \lambda^{\mbox{\tiny $(j)$}}(z)\:=\:\lambda^{\mbox{\tiny $(j)$}}+z\lambda^{\mbox{\tiny $(i)$}}.
\end{equation}

The conditions \eqref{zeroes-sum} strongly suggest a link among ``weights'' related to scattering amplitudes with different
number of external states. Such a link has not been established yet, even if the conditions \eqref{zeroes-sum} can be
solved for a number of examples \cite{Benincasa:2011kn}.

In the case of four-particle amplitudes, where we will label the particles by $i$, $j$, $k$, $m$, there are at most two 
helicity factors $\mathcal{H}_{n-1}^{(k)}$ in each of the last two equations of \eqref{zeroes-sum}, and they are given by
\begin{equation}\eqlabel{H4}
 \begin{split}
  &\mathcal{H}_{3}^{(k)}\:=\:\left(\frac{[k,m]}{[m,j]}\right)^{2(\delta+h_i)},\qquad
   \mathcal{H}_{3}^{(m)}\:=\:\left(-1\right)^{\delta}\left(\frac{[k,m]}{[j,k]}\right)^{2(\delta+h_i)},\\
  &\tilde{\mathcal{H}}_{3}^{(m)}\:=\:\left(\frac{\langle k,m\rangle}{\langle i,k\rangle}\right)^{2(\delta-h_j)},\qquad
   \tilde{\mathcal{H}}_{3}^{(k)}\:=\:\left(-1\right)^{\delta}
   \left(\frac{\langle k,m\rangle}{\langle m,i\rangle}\right)^{2(\delta-h_j)}.
 \end{split}
\end{equation}
The conditions \eqref{zeroes-sum} get heavily simplified

\begin{equation}\eqlabel{zeroes-sum-4}
 \begin{split}
  &\hspace{0.25cm}P_{ik}^2(z_0^{\mbox{\tiny $(l)$}})\:=\:\alpha^{\mbox{\tiny $(l)$}} P_{ij}^2,\qquad\qquad
    P_{im}^2(z_0^{\mbox{\tiny $(l)$}})\:=\:-(1+\alpha^{\mbox{\tiny $(l)$}}) P_{ij}^2,\\
  &1\:=\:\lim_{[i,j]\,\rightarrow\,0}\left[
    \left(-1\right)^{\nu+1+2h_k+\delta}\left(\prod_{l=1}^{\nu+1}\alpha^{\mbox{\tiny $(l)$}}\right)^{-1}
    \left(\frac{P_{ik}^2}{P_{ij}^2}\right)^{\nu-\delta-2h_i}+
   \right.\\
  &\hspace{1cm}+\left.
    (-1)^{2h_k}\left(\prod_{l=1}^{\nu+1}\left(1+\alpha^{\mbox{\tiny $(l)$}}\right)\right)^{-1}
    \left(\frac{P_{im}^2}{P_{ij}^2}\right)^{\nu-\delta-2h_i}
   \right],\\
  & 1\:=\:\lim_{\langle i,j\rangle\,\rightarrow\,0}\left[
    (-1)^{\nu+1+2h_m+2h_i+\delta}
    \left(\prod_{l=1}^{\nu+1}\alpha^{\mbox{\tiny $(l)$}}\right)^{-1}
    \left(\frac{P_{ik}^2}{P_{ij}^2}\right)^{\nu-\delta+2h_j}+
   \right.\\
  &\hspace{1cm}+\left.
    \left(-1\right)^{2h_i+2h_k}
    \left(\prod_{l=1}^{\nu+1}\left(1+\alpha^{\mbox{\tiny $(l)$}}\right)\right)^{-1}
    \left(\frac{P_{im}^2}{P_{ij}^2}\right)^{\nu-\delta+2h_j}
  \right],
 \end{split}
\end{equation}
where the last two relations require the parameter $\nu$ to be fixed to $\delta+2h_i$ and/or to $\delta-2h_j$ - we will 
further comment on this point later on. Once the large-$z$ parameter $\nu$ is fixed, the last two relations in 
\eqref{zeroes-sum-4} return a relation among the coefficients $\alpha^{\mbox{\tiny $(l)$}}$ which parametrise the
zeroes $z_0^{\mbox{\tiny $(l)$}}$:
\begin{equation}\eqlabel{zeroes-sum-4b}
 \begin{split}
  &1\:=\:
    \left(-1\right)^{2h_i+2h_k+1}\left(\prod_{l=1}^{\nu+1}\alpha^{\mbox{\tiny $(l)$}}\right)^{-1}+
    (-1)^{2h_k}\left(\prod_{l=1}^{\nu+1}\left(1+\alpha^{\mbox{\tiny $(l)$}}\right)\right)^{-1},\\
  &1\:=\:
    (-1)^{2h_k+1}\left(\prod_{l=1}^{\nu+1}\alpha^{\mbox{\tiny $(l)$}}\right)^{-1}+
    \left(-1\right)^{2h_i+2h_k}
    \left(\prod_{l=1}^{\nu+1}\left(1+\alpha^{\mbox{\tiny $(l)$}}\right)\right)^{-1}.
 \end{split}
\end{equation}
Notice that we assumed that both the $P_{ik}^2$ and $P_{im}^2$ channels are present in the BCFW decomposition. If only one 
of them is appearing, one has to drop the corresponding terms in \eqref{zeroes-sum-4} and \eqref{zeroes-sum-4b}

\subsection{Generalised On-Shell Recursion Relations and Constructibility}\label{subsec:GenConstr}

The existence of the generalised recursion relations \eqref{GenRR2} extends the notion of constructibility, in the sense 
that any non-trivial tree-level scattering amplitude can be expressed in terms of the lower-point ones.

The main consequence of the constructibility of a theory is that the tree-level scattering amplitudes can be determined just
from the knowledge of the three-particle amplitudes, which are fixed by momentum conservation and helicity scaling 
\cite{Benincasa:2007xk}, and by summing on a subset of the channels.

Interestingly, the notion of constructibility is related to the factorisation property of the amplitude in the collinear
channel involving the momenta which have been deformed \cite{Schuster:2008nh, Benincasa:2011kn}. Such a collinear 
singularity does not appear explicitly in the on-shell representation and it is realised as a soft limit of one of the 
deformed momenta \cite{Schuster:2008nh, Benincasa:2011kn}. More specifically, under the deformation \eqref{BCFWdef2} 
the collinear limit of interest is $P_{ij}^2\,\equiv\,\langle i,j\rangle[i,j]\,\rightarrow\,0$, which can be taken in two 
different ways, by sending to zero either the holomorphic or the anti-holomorphic inner product. Obviously, a given
amplitude does not necessarily need to factorise under both of these limits. 

As discussed in \cite{Schuster:2008nh, Benincasa:2011kn}, when the anti-holomorphic limit is taken on the on-shell 
representation, the only terms contributing are the ones with three-particle amplitudes containing particle-$i$, and the 
collinear limit translates into a limit in which the deformed momenta of particle-$i$ becomes soft (when the other limit is 
taken, the terms contributing are the ones with three-particle amplitudes containing particle-$j$, and the collinear limit 
translates into particle-$j$ becoming soft). 

Now, whether or not the standard BCFW representation is valid depends on 
whether or not this soft limit is able to produce a singularity which leads to the desired factorisation. The standard
BCFW representation fails when the soft limit by itself does not induce the correct factorisation, while in the 
generalised on-shell representation this soft singularity is produced by the ``weights'', and the amplitude factorises
properly. 

In the same fashion, whether or not the standard BCFW representation is valid depends on whether or not the one-parameter
family of amplitudes $M^{\mbox{\tiny $(i,j)$}}(z)$ generated by a certain BCFW-deformation vanishes as $z$ is taken to
infinity. 

In \cite{Benincasa:2011kn}, the analysis of this collinear limit on the generalised on-shell representation allows to 
univocally fix the complex-UV behaviour parameter $\nu$ in terms of the number of derivatives $\delta$ of the 
particular interaction and to the helicity of one of the deformed particles. Strictly speaking, this analysis holds
just for $\nu\,\ge\,0$ because it is used the fact that the ``weights'' in the representation \eqref{GenRR2}
depend explicitly on $\nu$ for $\nu\,\ge\,0$. However, as briefly described here, the validity of the standard
BCFW representation is intimately connected to the fact that the soft limit of (at least) one of the deformed particles
produce a singularity which allows the amplitude to factorise properly. As a consequence, the large-$z$ behaviour
of the particle is connected to such a soft limit. Furthermore, when the collinear limit is taken, the only terms
of the on-shell representation contributing show a three-particle amplitude containing the soft-particle, and, therefore,
the large-$z$ behaviour of the amplitude is connected to the soft limit on the three-particle amplitudes.

In some sense, there is a sort of ``duality'' between the large-$z$ behaviour of the amplitude, which is understood
as a hard particle propagating in a soft background \cite{ArkaniHamed:2008yf} and the soft-limit behaviour of 
three-particle amplitudes.

We are going to explicitly analyse the soft-limit behaviour for the three-particle amplitudes.

\subsubsection{Soft limits and Complex-UV Behaviour}\label{subsubsec:Soft}

In this section we argue that the degree of the pole/zero produced by a soft singularity in a three-particle amplitude
coincides with the behaviour of the amplitude in the large-$z$ limit. We first analyse generically the soft limits in a 
three-particle amplitude and later we will make contact with the expressions appearing in the collinear limit analysis in
\cite{Benincasa:2011kn} and with the known large-$z$ behaviours.

First, let us write here the expression for the three-particle amplitudes \cite{Benincasa:2007xk}:
\begin{equation}\eqlabel{3ptam}
 M_{3}\left(1^{\mbox{\tiny $h_{1}$}},2^{\mbox{\tiny $h_{2}$}},3^{\mbox{\tiny $h_{3}$}}\right)\:=\:
  \kappa^{\mbox{\tiny H}}_{\mbox{\tiny $1+h$}}
   \langle 1,2\rangle^{d_{3}}\langle 2,3\rangle^{d_{1}}\langle 3,1\rangle^{d_{2}}+
  \kappa^{\mbox{\tiny A}}_{\mbox{\tiny $1-h$}}
   [1,2]^{-d_{3}}[2,3]^{-d_{1}}[3,1]^{-d_{2}},
\end{equation}
where $d_{1}=h_{1}-h_{2}-h_{3}$, $d_{2}=h_{2}-h_{3}-h_{1}$, $d_{3}=h_{3}-h_{1}-h_{2}$, and 
$h\,=\,h_{1}+h_{2}+h_{3}$. The subscripts in the coupling constants indicate their dimension, while the superscript $H/A$ 
indicates the holomorphic/anti-holomorphic part of them amplitude. 
Notice that the amplitude \eqref{3ptam} has to go to zero as $\langle i,j\rangle$ and $[i,j]$
are both sent to zero, {\it i.e.} on the real sheet. This implies that if $d_{1}+d_{2}+d_{3}\equiv-h_{1}-
 h_{2}-h_{3}\,<\,0$, then the coupling constant $\kappa^{\mbox{\tiny H}}$ needs to be set to zero in order
to avoid infinities. Similarly, if $d_{1}+d_{2}+d_{3}\equiv-h_{1}- h_{2}-h_{3}\,>\,0$ then 
$\kappa^{\mbox{\tiny A}}$ needs to be set to zero. For $d_{1}+d_{2}+d_{3}\,=\,0$, both of the terms in
\eqref{3ptam} are allowed. It should also be noticed that $\delta\,=\,\left|d_{1}+d_{2}+d_{3}\right|$ provides the order
of the derivatives for the interaction (this can be understood through a simple dimensionality argument).
As pointed out in \cite{Benincasa:2007xk}, the expression \eqref{3ptam} for the three-particle amplitude is fully
non-perturbative.

For interactions with $\delta\,\equiv\,-h_1-h_2-h_3\,>\,0$, the three-particle amplitude is given just by
the holomorphic term. It is convenient to solve the relation connecting the helicities of the particles and the number of
derivatives $\delta$ for one of the helicities (let's say $h_3$) and substitute it into the expression for the three-particle
amplitude in such a way that the number of derivatives of the interaction becomes an explicit parameter:
\begin{equation}\eqlabel{3ptamHd}
 M_{3}\left(1^{\mbox{\tiny $h_{1}$}},2^{\mbox{\tiny $h_{2}$}},3^{\mbox{\tiny $h_{3}$}}\right)\:=\:
  \kappa\,\frac{\langle 2,3\rangle^{\delta+2h_1}\langle 3,1\rangle^{\delta+2h_2}}{\langle1,2\rangle^{\delta+2h_1+2h_2}}.
\end{equation}
We are going to analyse the amplitude \eqref{3ptamHd} when particle-$1$ becomes soft: 
$p^{\mbox{\tiny $(1)$}}\,\rightarrow\,0$. Thinking that the momentum of a massless particle in the complexified momentum
space is the direct product of two independent spinors, the soft limit can be taken in two different ways, {\it i.e.}
by sending either $\lambda^{\mbox{\tiny $(1)$}}$ or $\tilde{\lambda}^{\mbox{\tiny $(1)$}}$ to zero. 

For the limit $\lambda^{\mbox{\tiny $(1)$}}\,\rightarrow\,0$, let us choose 
$\lambda^{\mbox{\tiny $(1)$}}\,=\,\epsilon\,\eta$, so that the limit of interest is performed by taking the parameter
$\epsilon$ to zero. Moreover, recalling that for an amplitude such as 
\eqref{3ptamHd} momentum conservation implies that the anti-holomorphic spinors of the three particles are all
proportional to each other, we can set all of them to be equal. As a consequence, from momentum conservation, the
holomorphic spinors are related to each other through the relation 
$\lambda^{\mbox{\tiny $(3)$}}\,=\,-\epsilon\,\eta-\lambda^{\mbox{\tiny $(2)$}}\,$. Using these relations into 
\eqref{3ptamHd} we get
\begin{equation}\eqlabel{3ptHsoft}
 M_3(\epsilon)\:=\:\kappa\,(-1)^{2h_1+2h_2}\langle\eta,2\rangle^{\delta}\,\epsilon^{\delta},
\end{equation}
which vanishes as $\epsilon\,\rightarrow\,0$.

Let us now consider the anti-holomorphic limit $\tilde{\lambda}^{\mbox{\tiny $(1)$}}\,\rightarrow\,0$. Again,
the anti-holomorphic spinors are proportional to each other. We choose them to be
$\tilde{\lambda}^{\mbox{\tiny $(2)$}}\,=\,\tilde{\eta}\,=\,\tilde{\lambda}^{\mbox{\tiny $(3)$}}$ and
$\tilde{\lambda}^{\mbox{\tiny $(1)$}}\,=\,\epsilon\tilde{\eta}$, so that the soft limit is realised by taking $\epsilon$ to
zero. Through momentum conservation, the holomorphic spinors are related to each other as
$\lambda^{\mbox{\tiny $(3)$}}\,=\,-\epsilon\lambda^{\mbox{\tiny $(1)$}}-\lambda^{\mbox{\tiny $(2)$}}$. From these
relations among the spinors, the dependence of the amplitude on $\epsilon$ is
\begin{equation}\eqlabel{3ptHsoft2}
  M_3(\epsilon)\:=\:\kappa\,(-1)^{2h_1+2h_2}\langle1,2\rangle^{\delta}\,\epsilon^{\delta+2h_1},
\end{equation}
whose behaviour in the limit $\epsilon\,\rightarrow\,0$ depends on the sign of the exponent $\delta+2h_1$.

Similarly, the analysis of the anti-holomorphic three-particle amplitude -- which needs to be considered whenever
$h_1+h_2+h_3\,>\,0$ with $\delta\,=\,h_1+h_2+h_3$ -- leads to the following behaviours in the soft limits
\begin{equation}\eqlabel{3ptHsoft3}
 \begin{aligned}
  &M_3(\epsilon)\:=\:\kappa\,(-1)^{2h_1+2h_2}[1,2]^{\delta}\,\epsilon^{\delta-2h_1}\,, 
   &\lambda^{\mbox{\tiny $(1)$}}\,=\,\epsilon\eta,\\
  &M_3(\epsilon)\:=\:\kappa\,(-1)^{2h_1+2h_2}[\tilde{\eta},2]^{\delta}\,\epsilon^{\delta}\,, 
   &\tilde{\lambda}^{\mbox{\tiny $(1)$}}\,=\,\epsilon\tilde{\eta},
 \end{aligned}
\end{equation}
from which it is easy to infer that the amplitude vanishes as $\tilde{\lambda}^{\mbox{\tiny $(1)$}}\,\rightarrow\,0$,
while the behaviour of the amplitude in the limit $\lambda^{\mbox{\tiny $(1)$}}\,\rightarrow\,0$ depends on the
sign of the exponent $\delta-2h_1$.

Let us now make contact with the analysis of the collinear limit containing both deformed momenta done in Section 4.4 
of \cite{Benincasa:2011kn}. For this purpose, let us suppose that the deformed momenta belong to the particles labelled
by $1$ and $j$, for which the anti-holomorphic and the holomorphic spinors have respectively been shifted. When we analyse 
the collinear limit taken as $[1,j]\,\rightarrow\,0$, the only terms which might contain a singularity in this channel are
the ones containing a three-particle amplitude involving particle-$1$. For this amplitude, all the anti-holomorphic
spinors are proportional to each other and therefore it is expressed by the holomorphic term in \eqref{3ptam}. Furthermore,
the anti-holomorphic spinor of particle-$1$ turns out to be directly proportional to $[1,j]$ and hence it vanishes
in the limit $[1,j]\,\rightarrow\,0$. This case thus reduces to the one in \eqref{3ptHsoft2} with $\epsilon\,\sim\,[1,j]$. 

Similarly, the only terms which might contain a singularity as $\langle1,j\rangle\,\rightarrow\,0$
are the ones containing a three-particle amplitude involving particle-$j$. This amplitude is expressed in terms
of the anti-holomorphic spinors and the holomorphic spinor of particle-$j$ turns out to be directly proportional
to $\langle1,j\rangle$ so that it vanishes in this limit. Hence, this case reduces to the one in the first line of
\eqref{3ptHsoft3} with $\epsilon\,\sim\,\langle1,j\rangle$.

Therefore, the relevant soft-limit scalings are $\delta+2h_1$, in case the momentum $p^{\mbox{\tiny $(1)$}}$ becomes
soft through its anti-holomorphic spinor, and $\delta-2h_j$ in case the soft limit is taken by sending the holomorphic
spinor to zero.

As we mentioned in the previous subsection, the standard BCFW representation holds if the collinear singularity in
the $(1,j)$-channel appears as a soft singularity. If the amplitude admits just one factorisation limit: either
$[1,j]\,\rightarrow\,0$ or $\langle1,j\rangle\,\rightarrow\,0$, this requirement will be satisfied if and only if
$\delta+2h_1\,<\,0$ or $\delta-2h_j\,<\,0$ respectively. If both factorisation limits are allowed, the inequalities
just written down need to be satisfied simultaneously and their left-hand-sides need to coincide, relating the helicities
of the particles whose momenta have been deformed.

In the cases $\delta+2h_1\,\ge\,0$ and/or $\delta-2h_j\,\ge\,0$, the deformed particles still become soft in the relevant limit,
but by themselves they are not enough to provide with the correct pole. The introduction of the ``weights'', which depend 
on a subset of zeroes of the amplitude, enhances the soft limit to produce the correct pole. As we have previously seen,
these ``weights'' contain explicitly the large-$z$ parameter $\nu$, which the factorisation requirement fixes to be exactly
the soft-limit exponent(s) \cite{Benincasa:2011kn}.

At a conceptual level, it appears clear the connection between the soft exponents and the large-$z$ parameter $\nu$, both
for $\nu\,\ge\,0$ and for $\nu\,<\,0$. While the exact equivalence has been proven to be $\nu\,=\,\delta+2h_1$ and/or
$\nu\,=\,\delta-2h_j$ for $\nu\,\ge\,0$, so far we provided heuristic arguments for which this equivalence should hold also 
in the case $\nu\,<\,0$.

This issue can be overcome by substituting the large-$z$ criterion with the soft-limit one. More
precisely, with the generalised on-shell recursion relations \eqref{GenRR2} at hand, whether the ``weights'' are equal to 
one or not can be established by looking at the soft behaviour of the relevant three-particle amplitudes, and in the case where
the ``weights'' are required to be different than one, the soft behaviour fixes the parameter $\nu$.

Such a point of view is very suggestive because it brings (almost)\footnote{We are saying ``almost'' because, while we are
able to related the ``weights'' to the soft-behaviour of the three-particle amplitudes, we do not still find an exact 
expression which relates the ``weights'' of an amplitudes to the ones of the lower point ones.} everything back to 
three-particle level, even if at a practical level not much changed from the point of  view adopted in \cite{Benincasa:2011kn}. 

\subsubsection{Constructibility and Consistency Conditions}\label{subsubsec:ConsCond}

Constructibility gives us a very powerful consistency test for the existence of a non-trivial S-matrix
\cite{Benincasa:2007xk}. As we emphasised earlier, in constructible theories the physical amplitude should
be independent of the BCFW-deformation used to compute it. One can therefore consider the simplest 
non-trivial object, the four-particle amplitude, and compute it through two different BCFW-deformations.
Imposing that the two results coincide brings on non-trivial constraints on the S-matrix \cite{Benincasa:2007xk}
\begin{equation}\eqlabel{4ptTest}
 M_{4}^{\mbox{\tiny $(i,j)$}}(0)\:=\:M_{4}^{\mbox{\tiny $(i,k)$}}(0).
\end{equation}
Through this ``four-particle test'', one can rediscover for example the Jacobi-identity in Yang-Mills and 
$\mathcal{N}=1$ Supergravity, in which both the gauge symmetry and the supersymmetry emerge from the consistency of the 
theory rather than being postulated a priori.

This suggests that interactive theories may have a much simpler formulation based on a minimal amount of
assumptions (listed at the beginning of Section \ref{sec:BCFWrev}), which lead to a single type of building block
(the three-particle amplitude \eqref{3ptam}).

In what follows, with the generalised notion of constructibility discussed in Section \ref{subsubsec:Soft} and the 
generalised on-shell recursion relations \eqref{GenRR2} at hand, we will use the consistency criterion \eqref{4ptTest} to
further explore the space of consistent interacting theories of massless particles. The minimum goal is to rediscover
all the known theories which were missed by applying the criterion \eqref{4ptTest} with the notion of constructibility
provided by the standard BCFW recursion relations. More interestingly, we can aim at finding a signature of the possible
existence of non-trivial high-spin interactions or a signature of the breaking down of one of the fundamental properties
when particles with spin higher than two are considered.

\section{Four-Particle Amplitudes}
\label{sec:4pt2}

In this section we discuss the program, outlined earlier, for the particular case of the four-particle 
amplitudes. Considering the momentum-space deformation
\begin{equation}\eqlabel{BCFWx}
 \tilde{\lambda}^{\mbox{\tiny $(i)$}}(z)\:=\:\tilde{\lambda}^{\mbox{\tiny $(i)$}}-z\tilde{\lambda}^{\mbox{\tiny $(j)$}},
  \qquad
 \lambda^{\mbox{\tiny $(j)$}}(z)\:=\:\lambda^{\mbox{\tiny $(j)$}}+z\lambda^{\mbox{\tiny $(i)$}},
\end{equation}
the conditions \eqref{zeroes-sum-4} reduce to 
\begin{equation}\eqlabel{zeroes-4}
 \prod_{r\,=\,1}^{N_{P}^{\mbox{\tiny fin}}}P^2_{i s_r}(z_0)\:=\:
 (-1)^{N_P^{\mbox{\tiny fin}}}\left(P^2_{ij}\right)^{N_P^{\mbox{\tiny fin}}},
\end{equation}
where $s_r$ can be either $k$ or $m$ which label the two particles other than $i$ and $j$, and $N_P^{\mbox{\tiny fin}}$ is the
number of poles at finite location, which can only be one or two for four-particle amplitudes.

Some comments are now in order. Strictly speaking, there are cases in which the one-parameter family of amplitudes
$M_4^{\mbox{\tiny $(i,j)$}}(z)$ might not show any pole at finite location. In such cases, the full amplitude would
correspond to the contribution $\mathcal{C}_4^{\mbox{\tiny $(i,j)$}}$ from the singularity at infinity and the method
outlined in Section \ref{subsec:GenRR} breaks down. In any case, it is always possible to choose a deformation of the 
momentum space such that $M_4^{\mbox{\tiny $(i,j)$}}(z)$ shows at least a pole. However, also in this case there is
an issue arising. Our conditions on the zeroes have been deduced by looking at the factorisation properties of the 
amplitude in the $(i,j)$-channel, while the amplitudes of these theories do not possess such a factorisation channel.
We will further discuss this issue in the next sections.

Furthermore, a given four-particle amplitudes does not necessarily factorise in both limits 
$\langle i,j\rangle\,\longrightarrow\,0$ and 
$[i,j]\,\longrightarrow\,0$\footnote{Notice that the limits $\langle i,j\rangle\,\longrightarrow\,0$ and
$[i,j]\,\longrightarrow\,0$ are equivalent to $[k,l]\,\longrightarrow\,0$ and $\langle k,l\rangle\,\longrightarrow\,0$ 
respectively.}. When the two factorisations are allowed, the behaviour of $M_4^{\mbox{\tiny $(i,j)$}}(z)$ as $z$ is taken 
to infinity is fixed simultaneously by both of the last two conditions in \eqref{zeroes-sum-4}
\begin{equation}\eqlabel{LargeZnu}
 \nu\:=\:\delta+2h_i\,=\,\delta-2h_j, \qquad (\nu\,\ge\,0)
\end{equation}
which implies a relation between the helicities of the particles whose momenta have been deformed: $h_j\,=\,-h_i$.
It is instructive to make a systematic analysis of the complex-UV behaviour, or, as from the discussion in Section
\ref{subsubsec:Soft}, of the soft-limit behaviour, in the space of all the possible consistent four-particle amplitudes.

First, a classification of the theories
can be done through the dimensionality of the three-particle coupling constant, {\it i.e.} in a Lagrangian language, through
the number of derivatives of the three-particle interaction:

\begin{enumerate}
 \item $[\kappa]\,=\,1-s$ ($s$-derivative interactions: $\delta\,=\,s$). 
       This class contains two sub-classes of theories: self-interacting particle of spin-$s$ with three-particle amplitudes 
       having, as possible helicity configurations, $(\mp s, \mp s, \pm s)$; and spin-$s$-spin-$s'$ interactions, whose 
       three-particle amplitudes may have, as possible helicity configurations, $(-s',+s',\mp s)$;
 \item $[\kappa]\,=\,1-3s$ ($3s$-derivative interactions: $\delta\,=\,3s$). It contains self-interacting particle
       of spin-$s$, whose three-particle amplitudes admit the helicity configuration $(\mp s, \mp s, \mp s)$;
 \item $[\kappa]\,=\,1-(2s'+s)$ ($(2s'+s)$-derivative interactions: $\delta\,=\,2s'+s$). 
       It is characterised by three-particle amplitudes whose helicity structure may be $(\mp s', \mp s', \mp s)$;
 \item $[\kappa]\,=\,1-|2s'-s|$ ($|2s'-s|$-derivative interactions: $\delta\,=\,|2s'-s|$). 
       It is characterised by three-particle amplitudes whose helicity configuration may be 
       $(\mp s', \mp s', \pm s)$. Depending on whether $s'$ is less or greater than
       $2s$, the three-particle amplitude with a given helicity configuration (between the two allowed) can be represented
       by the holomorphic term in \eqref{3ptam} in one case or the anti-holomorphic one in the other case. For
       $s\,=\,2s'$, the theories have $0$-derivative interactions, and both the holomorphic and the anti-holomorphic pieces are present in the three-particle amplitude.
\end{enumerate}

Second, for each of the above classes of theories it is possible to fix the conditions on the zeroes and thus the
``weights'' in the generalised on-shell representation. 

\subsection{Interactions with $s$-derivatives: Self-interaction of spin-$s$ particles}\label{subsec:s}

Let us start to illustrate our analysis by taking into consideration the scattering of particles of spin $s$ whose coupling
has dimension $[\kappa]\,=\,1-s$. From \eqref{3ptam}, it turns out that such theories are characterised by two possible 
helicity configurations for the three-particle amplitudes if $s\,\neq\,0$:
\begin{equation}\eqlabel{Amplds}
 \begin{split}
  &M_{3}\left(1^{\mbox{\tiny $-s$}},\,2^{\mbox{\tiny $-s$}},3^{\mbox{\tiny $+s$}}\right)\:=\:
   \kappa\,\varepsilon_{\mbox{\tiny $a_1 a_2 a_3$}}
    \left(\frac{\langle1,2\rangle^3}{\langle2,3\rangle\langle3,1\rangle}\right)^{s},
  \\
  &M_{3}\left(1^{\mbox{\tiny $+s$}},\,2^{\mbox{\tiny $+s$}},3^{\mbox{\tiny $-s$}}\right)\:=\:
   \kappa\,\varepsilon_{\mbox{\tiny $a_1 a_2 a_3$}}\left(\frac{[1,2]^3}{[2,3][3,1]}\right)^{s},
 \end{split}
\end{equation}
where $\varepsilon_{\mbox{\tiny $a_1 a_2 a_3$}}$ are structure constants related to possible internal quantum numbers.
In the case that the theory does not show any internal symmetry, these structure constants can be set to one. It is easy to see that 
the structure constant needs to be completely symmetric in their indices for even spin $s$, while completely anti-symmetric 
for odd spin $s$.

In the case one is dealing with a scalar theory, the three-particle amplitude is given by a coupling constant, which is given
by the sum of the holomorphic and anti-holomorphic coupling constants in \eqref{3ptam}. 

The three-point helicity configurations admitted in this class of theories allow to have just one class of non-trivial
four-point amplitudes which is characterised by having two particles with negative helicity and two with positive helicity.

\begin{table}
  \centering
  \begin{tabular}{|c||*{2}{c|}}\hline
   \backslashbox{$h_i$}{$h_j$} & $-s$ & $+s$ \\ \hline
   $ -s $ & \backslashbox{\color{red} $-s$}{X} & {\color{red} $-s$} \\ \hline
   $ +s $ & {\color{red} $3s$} & \backslashbox{X}{\color{red} $-s$} \\ \hline
  \end{tabular}\hfill
  \caption{Complex-UV behaviour $\nu$ for self-interacting particles of spin-$s$. In this table, the complex-UV behaviour
           $\nu$ is shown in red as a function of the helicities $h_i$ and $h_j$ of the particles whose momenta have been 
           deformed.  The cells containing a single value for $\nu$ correspond to the cases where the amplitude factorises 
           both in the $\langle i,j\rangle\,\rightarrow\,0$ and $[i,j]\,\rightarrow\,0$ limits. In the other cells, the 
           value in the lower (upper) triangle corresponds to the case in which the amplitude factorises in the 
           $[i,j]\,\rightarrow\,0$ ($\langle i,j\rangle\,\rightarrow\,0$) limit.
           }
  \label{UVds2}
\end{table}

Let us consider the following two-particle deformation:
\begin{equation}\eqlabel{BCFWs}
 \tilde{\lambda}^{\mbox{\tiny $(1)$}}(z)\:=\:\tilde{\lambda}^{\mbox{\tiny $(1)$}}-z\tilde{\lambda}^{\mbox{\tiny $(2)$}},
 \qquad
 \lambda^{\mbox{\tiny $(2)$}}(z)\:=\:\lambda^{\mbox{\tiny $(2)$}}+z\lambda^{\mbox{\tiny $(1)$}},
\end{equation}
where for the moment the particles are kept with arbitrary helicities. First of all, let us discuss the complex-UV limit.
A complete analysis of the behaviour of the amplitude at infinity as a function of the helicities of the particles is
displayed in Table \ref{UVds2}, where $h_i$ and $h_j$ need to be identified with $h_1$ and $h_2$. 
Let us comment on this more extensively. The first feature to notice is that the choices 
$(h_1,\,h_2)\,=\,(-s,\,+s)$ and $(h_1,\,h_2)\,=\,(+s,\,-s)$
lead respectively to the behaviours $z^{-s}$ and $z^{3s}$. This is in agreement with the known results for Yang-Mills
and Gravity under the standard BCFW-deformation ($\sim\left.z^{-s}\right|_{s=1,2}$) and the ``wrong'' one 
($\sim\left.z^{3s}\right|_{s=1,2}$). The other two choices for the helicities of the deformed particles seem to show two 
possible values for the parameter $\nu$, depending on how the collinear limit is taken. This puzzle is quickly resolved
by noticing that the amplitude factorises under just one of the two ways in which the limit $P_{12}^2\,\rightarrow\,0$ can
be realized. It is easy to understand also which limit is allowed by just looking at the helicity configuration of the
three-particle amplitude in which the amplitude would eventually factorise. Let us consider for instance $(h_1,\,h_2)\,=\,(-s,\,-s)$.
Under the limit $\langle 1,\,2\rangle\,\longrightarrow\,0$, the amplitude $M_4(1^{-s},2^{-s},3^{+s},4^{+s})$ would factorise
into 
\begin{equation}\eqlabel{facts1}
 \lim_{\langle 1,2\rangle\,\longrightarrow\,0}P_{12}^2 M_4 \:=\:
 M_3\left(1^{-s},2^{-s},-P_{12}^{h_{12}}\right)M_3\left(P_{12}^{-h_{12}},3^{+s},4^{+s}\right).
\end{equation}
The class of theories we are discussing admits just the three-particle amplitudes \eqref{Amplds} and therefore the
helicity $h_{12}$ is fixed to be $h_{12}\,=\,+s$. It is clear from \eqref{Amplds} that in the limit $\langle 1,\,2\rangle\,\longrightarrow\,0$, the sub-amplitude 
$M_3\left(1^{-s},2^{-s},-P_{12}^{h_{12}}\right)$ vanishes.

The right-hand-side of \eqref{facts1} represents as well the factorisation of the four-particle amplitude in the limit
$[1,2]\,\longrightarrow\,0$. In this case, none of the two three-particle amplitudes vanishes and the amplitude does
factorise in this limit. 

Therefore, the complex-UV behaviour with this choice of the helicities is $\sim z^{-s}$ as in the the lower triangle in the 
first cell of table \ref{UVds2}. Moreover, the limits $\langle 1,2\rangle\,\longrightarrow\,0$ and 
$[1,2]\,\longrightarrow\,0$ are equivalent to $[3,4]\,\longrightarrow\,0$ and 
$\langle3,4\rangle\,\longrightarrow\,0$ respectively. The previous analysis thus implies that the amplitude factorises just 
under the limit $\langle3,4\rangle\,\longrightarrow\,0$. As a consequence, picking the choice $(+s,\,+s)$ for the helicities
of the deformed particles, the complex-UV behaviour is $\sim z^{-s}$, as indicated in the upper triangle in last cell of
table \ref{UVds2}. 

In the case of a scalar theory, all the factorisation limits are allowed and the amplitude behaves as a constant as $z$
is taken to infinity.

Having established generically the complex-UV limit, we now focus on the computation of the four-point amplitude, that will allow us to run the four-particle test. As
a first helicity choice for the particle whose momenta we deform as in \eqref{BCFWs} we pick $(h_1,\,h_2)\,=\,(-s,\,+s)$.
With such a choice, the fall-off of the amplitude as $z$ is taken to infinity is $z^{-s}$, and therefore for $s\,\neq\,0$
the amplitude admits the standard BCFW representation. However, as it was shown in \cite{Benincasa:2007xk}, the only
consistent theories admitting such a representation are given by $s\,=\,1$ with internal quantum numbers and $s\,=\,2$. 
In the case of spin-$1$, the equality was holding if and only if the structure constants were satisfying the Jacobi
identity, while for the spin-$2$ particles the algebra is reducible and leads to several self-interacting spin-$2$ particles
which do not interact with each other. 

For completeness, let us analyse the only missed case, the scalar case. Picking the deformation \eqref{BCFWs} and
using the fact that the three-particle amplitude is just the coupling constant 
$\hat{\kappa}\,=\,\kappa_{\mbox{\tiny H}}+\kappa_{\mbox{\tiny A}}$, the on-shell representation for the four-particle
amplitude turns out to be
\begin{equation}\eqlabel{M4s0}
 M_{4}^{(1,2)}(0)\:=\:\hat{\kappa}^2\frac{f_{13}^{\mbox{\tiny $(0,4)$}}}{P_{13}^2}+
  \hat{\kappa}^2\frac{f_{14}^{\mbox{\tiny $(0,4)$}}}{P_{14}^2},
\end{equation}
where the notation on the left-hand-side is meant to stress the fact that this expression has been obtained by deforming
the momenta of the particles labelled by $1$ and $2$. The amplitude behaves as a constant at infinity, which implies that
we need the knowledge of just one zero in order to fix the ``weights'' $f_{1k}^{\mbox{\tiny $(0,4)$}}$ in \eqref{M4s0}.
Using the conditions \eqref{zeroes-sum-4} on the zeroes, it can be written as follows
\begin{equation}\eqlabel{M4s0b}
  M_{4}^{(1,2)}(0)\:=\:\hat{\kappa}^2\left(\frac{1}{P_{13}^2}+\frac{1}{P_{14}^2}-\frac{1}{\alpha(1+\alpha)P_{12}^2}\right).
\end{equation}
The condition \eqref{zeroes-4} further implies that the coefficient $\alpha$ needs to satisfy the equation
$\alpha\left(1+\alpha\right)\,=\,-1$, where the left-hand-side is exactly the form in which $\alpha$ enters in \eqref{M4s0b}.
Therefore, the final answer from the $(1,2)$-deformation is
\begin{equation}\eqlabel{M4s0c}
 M_{4}^{(1,2)}(0)\:=\:-\hat{\kappa}^2\frac{{\sf st} + {\sf tu} + {\sf us}}{{\sf s t u}},
\end{equation}
where the Mandelstam variables ${\sf s}\,\overset{\mbox{\tiny def}}{=} P_{12}^2$, 
${\sf t}\,\overset{\mbox{\tiny def}}{=} P_{14}^2$ and ${\sf u}\,\overset{\mbox{\tiny def}}{=} P_{13}^2$
have been introduced. It is easy to notice that the contribution from the singularity at infinity, provided by the last
term in \eqref{M4s0b}, contains the pole in the ${\sf s}$-channel which could not be reproduced by the residues
of the poles at finite positions.

One can try to perform again this computation by deforming the momenta of the particles $1$ and $4$. Notice that the result 
of this can be just obtained from \eqref{M4s0b} and \eqref{M4s0c} by the label exchange $2\,\longleftrightarrow\,4$. 
However, it is easy to see that the expression in \eqref{M4s0c} is invariant under such a label exchange and, therefore, the
$(1,4)$-deformation returns the same result. Therefore, the scalar theory passes the tree-level consistency check, as it
should, and eq \eqref{M4s0c} is the known result from the $\lambda\phi^3$-theory.

Furthermore, one can think of considering several species of scalars by introducing internal quantum numbers as in 
\eqref{Amplds}. For this case, as for all the cases of even spin, the structure constants are completely symmetric. 
Imposing the four-particle test, the consistency requirement implies an algebra structure similar to the one
found for several species of spin-$2$ particles in \cite{Benincasa:2007xk}, which is reducible. As a consequence,
this theory would reduce to a set of self-interacting scalars which do not interact among them.

Thus, the generalised on-shell representation allows to obtain all the known self-interacting theories and discard the 
existence of higher-spin self-interactions for this class of theories.

\subsection{Interactions with $s$-derivatives: spin-$s$/spin-$s'$}\label{subsec:ssp}

Consider interactions with the same dimension of the coupling $[\kappa]\,=\,1-s$, as in Section \ref{subsec:s}, but now
allowing also for particles of spin $s$ and $s'$. These interactions are defined through the three-particle amplitudes 
written in \ref{Amplds}, which describe the self-interaction of the particle of spin-$s$, and two further three-particle 
amplitudes describing the spin-$s$/spin-$s'$ interaction:
\begin{equation}\eqlabel{Ampldssp}
 \begin{split}
  &M_{3}(1^{-s'},2^{+s'}, 3^{-s})\:=\:\kappa'\varepsilon_{b_1 b_2 a_3}
    \frac{\langle 3,1\rangle^{s+2s'}}{\langle 1,2\rangle^{s}\langle 2,3\rangle^{2s'-s}},\\
  &M_{3}(1^{-s'},2^{+s'}, 3^{+s})\:=\:\kappa'\varepsilon_{b_1 b_2 a_3}
    \frac{[2,3]^{s+2s'}}{[3,1]^{2s'-s}[1,2]^{s}},
 \end{split}
\end{equation}
where $\varepsilon_{b_1 b_2 a_3}$ is an eventual structure constant whose indices $b$ are referred to the particles
of spin $s'$, while the index $a$ refers to the particle of spin-$s$. Moreover we keep the spin-$s$/spin-$s'$ coupling
constant to be different from the spin $s$ self-interaction one and an eventual relation should emerge from consistency 
requirements. Finally, the spin-$s'$ self-interaction is not allowed because we are focusing on interactions with the
same coupling constant dimensions ({\it i.e.} with a fixed number of derivatives in the interactions) and this would fix
$s'$ to be equal to $s$. It is certainly interesting to explore theories whose interactions have different couplings, but
we leave that to future work.

The four-particle analysis involves two types of amplitudes:
\begin{equation}\eqlabel{M4ssp}
 M_{4}(1^{-s},2^{+s},3^{-s'},4^{s'}),\qquad  M_{4}(1^{-s'},2^{+s'},3^{-s'},4^{+s'}),
\end{equation}
which are characterised respectively by three and two factorisation channels (in the second amplitude the ${\sf u}$-channel
is not permitted), and under a two-particle deformation they show two and one pole in $z$ respectively.

As a first step, let us analyse in detail the complex-UV behaviour of the amplitudes in \eqref{M4ssp} under the $(1,2)$-deformation
\eqref{BCFWs}. Such an analysis can be done in great generality if we do not fix the helicities and leave them to be
generically $(h_1,\,h_2)$. The complex-UV exponents of the amplitudes under all the possible two-particle deformations
are listed in Table \ref{UVds2b} (as in Section \ref{subsec:s}, one needs the identification 
$(h_i,\,h_j)\,\equiv\,(h_1,\,h_2)$).

\begin{table}[t]
 \centering
 \begin{tabular}{|c||*{4}{c|}}\hline
  \backslashbox{$h_i$}{$h_j$} & $-s$ & $+s$ & $-s'$ & $+s'$ \\\hline
  $-s$ & \backslashbox{\color{red} $-s$}{X} & {\color{red} $-s$} & \backslashbox{\color{red} $-s$}{X} & \backslashbox{\color{red} $-s$}{X}\\\hline
  $+s$ & {\color{red} $3s$} & \backslashbox{X}{\color{red} $\phantom{+s}-s$} & \backslashbox{X}{\color{red} $s+2s'$} & \backslashbox{X}{\color{red} $s-2s'$}\\\hline
  $-s'$ & \backslashbox{\color{red} $s-2s'$}{X} & \backslashbox{X}{\color{red} $-s$} & \backslashbox{X}{X} & {\color{red} $s-2s'$}\\\hline
  $+s'$ & \backslashbox{\color{red} $s+2s'$}{X} & \backslashbox{X}{\color{red} $-s$} & {\color{red} $s+2s'$} & \backslashbox{X}{X}\\\hline
 \end{tabular}
 \caption{Complex-UV behaviour $\nu$ for spin-$s$/spin-$s'$ interactions. Here, the complex-UV behaviour $\nu$ is shown as a 
          function of the helicities $h_i$ and $h_j$ in the case of particles with different spin. The values in red 
          indicate the actual complex-UV behaviour for each given choice of $(h_i,\,h_j)$. For the cells divided into two
          sub-cells, the upper and lower triangles are related to the factorisation in the holomorphic and anti-holomorphic 
          limits respectively. The red value therefore indicates as well which one in between these two limits lead to a 
          factorisation of the amplitude, while the ``X'' indicates that the amplitude does not have such a factorisation
          limit. The cells where both triangles show ``X'' indicate that none of these two limits is allowed.}
 \label{UVds2b}
\end{table}

The behaviour of the amplitudes when the helicities of particle-$1$ and particle-$2$ are chosen to be 
$(h_1,\,h_2)\,=\,(\mp s,\,\pm s)$ is the same as in the self-interacting case of the previous section for both of the 
two amplitudes under analysis. As before, when $h_1$ and $h_2$ are the same or they correspond to particles
of different spin, just one between the holomorphic and anti-holomorphic factorisation in the $(1,2)$-channel is allowed.
Following the argument of the previous section, it is easy to see which one occurs and therefore which one of them fixes the complex-UV behaviour. In Table \ref{UVds2b}, the values in 
red represent the complex-UV behaviour under a given assignment for the helicities of the particles whose momenta have been
deformed. Under some particular choices, namely $(h_1,\,h_2)\,=\,\left\{(-s',\,-s'),\,(+s',\,+s')\right\}$, neither
the holomorphic nor the anti-holomorphic factorisation in the $(1,2)$-channel are allowed and therefore, in  
principle, the analysis of the factorisation properties of the amplitudes to fix the large-$z$ parameter $\nu$ and the
Mandelstam variables when the S-matrix becomes trivial, seems to break down. We will comment on this later. For the moment,
we can continue our discussion on the four-particle amplitudes in this class of theories given that, as it is manifest from 
Table \ref{UVds2b}, it is always possible to choose an assignment for the helicities $(h_1,\,h_2)$ such that at least 
one of the factorisation limits in the $(1,2)$-channel holds and, as a consequence, the conditions \eqref{zeroes-sum-4} 
and \eqref{zeroes-4} are rigorously valid.

Notice that for the four-point amplitude $M_{4}(1^{-s},2^{+s},3^{-s'},4^{+s'})$, with $s\,\neq\,0$, it is always possible
to choose a deformation, namely the one defined by eq \eqref{BCFWs}, for which the amplitude vanishes as $\,\sim\,z^{-s}$.
As far as the amplitude $M_{4}(1^{-s'},2^{+s'},3^{-s'},4^{+s'})$ is concerned, the deformation \eqref{BCFWs} induces the
large-$z$ behaviour $M_4^{\mbox{\tiny $(1,2)$}}(z)\,\sim\,z^{s-2s'}$ and, therefore, the standard BCFW recursive relation
is valid if and only if $s\,<\,2s'$, while for $s\,\ge\,2s'$ the amplitude shows a recursive structure through the 
generalised on-shell formula \eqref{GenRR2}.

Let us start the detailed analysis of the possible constraints on the four-particle amplitudes by looking at the amplitude of
four external states with spin $s'$: $M_{4}(1^{-s'},2^{+s'},3^{-s'},4^{+s'})$. Under the deformation \eqref{BCFWs}, the
amplitude is given by
\begin{equation}\eqlabel{M4spsps12}
 M_{4}^{\mbox{\tiny $(1,2)$}}(1^{-s'},2^{+s'},3^{-s'},4^{+s'})\:=\:(\kappa')^2 (-1)^s \frac{f_{14}}{\sf t}
  \frac{\langle1,3\rangle^{2s'}[4,2]^{2s'}}{{\sf s}^{2s'-s}}.
\end{equation}
Similarly, under the $(1,4)$-deformation
\begin{equation}\eqlabel{BCFWs2}
 \tilde{\lambda}^{\mbox{\tiny $(1)$}}(z)\:=\:\tilde{\lambda}^{\mbox{\tiny $(1)$}}-z\tilde{\lambda}^{\mbox{\tiny $(4)$}},
 \quad
 \lambda^{\mbox{\tiny $(4)$}}\:=\:\lambda^{\mbox{\tiny $(4)$}}+z\lambda^{\mbox{\tiny $(1)$}},
\end{equation}
we get
\begin{equation}\eqlabel{M4spsps14}
 M_{4}^{\mbox{\tiny $(1,4)$}}(1^{-s'},2^{+s'},3^{-s'},4^{+s'})\:=\:(\kappa')^2 (-1)^s \frac{f_{12}}{\sf s}
  \frac{\langle1,3\rangle^{2s'}[4,2]^{2s'}}{{\sf t}^{2s'-s}}.
\end{equation}
Imposing the four-particle test
\begin{equation}\eqlabel{4ptTspsps}
 \frac{M_{4}^{\mbox{\tiny $(1,4)$}}}{M_{4}^{\mbox{\tiny $(1,2)$}}}\,=\,1\,=\,\frac{f_{12}}{f_{14}}
  \left(\frac{\sf s}{\sf t}\right)^{2s'-s-1}.
\end{equation}
As we mentioned earlier, for $s\,<\,2s'$ the one-parameter families of amplitudes generated by \eqref{BCFWs} and 
\eqref{BCFWs2} vanish as $z$ is taken to infinity, and the ``weights'' appearing in the generalised on-shell representation
are $1$. As a consequence, the consistency relation \eqref{4ptTspsps} is satisfied if and only if $s\,=\,2s'-1$.

In the case $s\,\ge\,2s'$, the ``weights'' $f_{14}$ and $f_{12}$ are fixed through the condition \eqref{zeroes-4}:
\begin{equation}\eqlabel{M4wf}
 \begin{split}
  &P_{14}^{2}(z_0^{\mbox{\tiny $(l)$}})\:=\:-P_{12}^2\qquad\Rightarrow\qquad 
   f_{14}\:=\:(-1)^{s-2s'+1}\left(\frac{\sf u}{\sf s}\right)^{s-2s'+1};\\
  &P_{12}^{2}(z_0^{\mbox{\tiny $(l)$}})\:=\:-P_{14}^2\qquad\Rightarrow\qquad 
   f_{12}\:=\:(-1)^{s-2s'+1}\left(\frac{\sf u}{\sf t}\right)^{s-2s'+1},
 \end{split}
\end{equation}
and the consistency condition \eqref{4ptTspsps} turns out to be identically satisfied.

Let us now focus on the amplitude $M_{4}(1^{-s},2^{+s},3^{-s'},4^{+s'})$ and let us compute it through the deformations
$(1,2)$ and $(1,4)$. First of all, from Table \ref{UVds2b} both one-parameter families of amplitudes generated by
such deformations behave as $\sim\,z^{-s}$ as $z$ is taken to infinity.
\begin{equation}\eqlabel{M4ssspsp12}
 \begin{split}
  &M_{4}^{\mbox{\tiny $(1,2)$}}\:=\:\left(\kappa'\right)^2 \frac{\langle1,3\rangle^s\langle1,4\rangle^s[4,2]^{2s'+s}}{
    [3,2]^{2s'-s}{\sf s}^s}\left(\frac{f_{13}^{\mbox{\tiny $(1,2)$}}}{\sf u}+\frac{f_{14}^{\mbox{\tiny $(1,2)$}}}{\sf t}
    \right),\\
  &M_{4}^{\mbox{\tiny $(1,4)$}}\:=\:\kappa'\frac{\langle1,3\rangle^s[4,2]^{2s'+s}}{[1,4]^s[3,2]^{2s'-s}}
    \left(\kappa\frac{f_{12}^{\mbox{\tiny $(1,4)$}}}{\sf s}+\kappa'\frac{f_{13}^{\mbox{\tiny $(1,4)$}}}{\sf u}\right),
 \end{split}
\end{equation}
where, differently from the previous notation, we are using $f_{1k}^{\mbox{\tiny $(1,j)$}}$ to relate the ``weights''
to the correspondent deformation.

As long as $s\,\neq\,0$, the ``weights'' are $1$ and the consistency condition reads
\begin{equation}\eqlabel{4ptTssspsp}
 \frac{M_{4}^{\mbox{\tiny $(1,4)$}}}{M_{4}^{\mbox{\tiny $(1,2)$}}}\,=\,1\,=\,
 (-1)^s\left(\frac{\sf s}{\sf t}\right)^{s-2}\left[1-\frac{\sf u}{\sf t}\left(\frac{\kappa}{\kappa'}-1\right)\right],
\end{equation}
which is satisfied if and only if $s\,=\,2$ and $\kappa'\,=\,\kappa$, or $s\,=\,1$ and $\kappa\,=\,0$.

For $s\,=\,0$, the ``weights'' are no longer one and the consistency condition becomes
\begin{equation}\eqlabel{4ptTssspsp2}
 \frac{M_{4}^{\mbox{\tiny $(1,4)$}}}{M_{4}^{\mbox{\tiny $(1,2)$}}}\,=\,1\,=\,
 \frac{\kappa}{\kappa'}-\left(\frac{\kappa}{\kappa'}-1\right)\frac{{\sf st} + \alpha{\sf us}}{{\sf st}+{\sf tu}+{\sf us}},
\end{equation}
where $\alpha$ is the parameter characterising the zeroes in \eqref{zeroes-sum-4}. The four-particle test turns out to 
be satisfied if and only if $\kappa'\,=\,\kappa$.

Summarising, the four-particle test on the amplitude with two external states of spin-$s$ and two with spin-$s'$ 
either sets the coupling constant $\kappa\,=\,0$ and the interaction mediator to have spin-$1$, or forces
the coupling constants $\kappa$ and $\kappa'$ to be identical and the interaction mediator to have spin-$0$ or
spin-$2$. When the test is instead applied to the amplitude with four external states of spin-$s'$, we either obtain
an exact relation between spin-$s$ and spin-$s'$, {\it i.e.} $s\,=\,2s'-1$, if $s\,<\,2s'$, or no constraint at all
for $s\,\ge\,2s'$. All together, these relations strongly constrain the types of theories we can have in this class.
Specifically, if $s\,=\,2$, we can only have $s'\,=\,3/2$ for $s\,<\,2s'$ when the standard BCFW relations hold, as it was
already seen in \cite{Benincasa:2007xk}, and $s'\,\le\,1$ for $s\,\ge\,2s'$.

For $s\,=\,1$, the condition for $s<2s'$ required by the consistency of the amplitude with all the external states of 
spin-$s$ is never fulfilled. For $s\,\ge\,2s'$, instead, we rediscover the couplings between spin-$1$ and fermions/
scalars. Notice that the self-interaction coupling $\kappa$ for $s\,=\,1$ needs to be zero, which implies that the
spin-$1$ mediator is actually a photon. Therefore, we have been rediscovering QED and scalar-QED.

Finally, if instead $s\,=\,0$, the only theory admitted has $s'\,=\,1/2$.  

Hence, the generalised on-shell recursion relations \eqref{GenRR2} allow us to rediscover not only $\mathcal{N}=1$ 
Supergravity, but also Einstein-Maxwell, Fermion-Gravity, Scalar-Gravity, QED, Yukawa theories and all the known
theories.

In the previous calculations we set the structure constants $\varepsilon_{b_1 b_2 a_3}$ to $1$. If instead we 
allow the theory to have an internal symmetry, it is easy to see what follows. From the analysis of the amplitude with four 
external states of spin-$s'$, the consistency condition \eqref{4ptTspsps} becomes
\begin{equation}\eqlabel{4ptTspspsC}
 \sum_{a_P}\varepsilon_{b_1 b_4 a_P}\varepsilon_{a_P b_3 b_2}\:=\:
 \sum_{a_P}\varepsilon_{b_1 b_2 a_P}\varepsilon_{a_P b_3 b_4}\frac{f_{12}}{f_{14}}\left(\frac{\sf s}{\sf t}\right)^{2s'-s-1}
\end{equation}
where the index $a$ in the structure constants is related to the spin-$s$ particles, while the index $b$ is related to
the spin-$s'$ ones. The conditions on the spins do not change with respect to the case where the theory was not endowed with
an internal symmetry: the theories which satisfy the standard BCFW representation are characterised by $s\,=\,2s'-1$,
while the others must have $s\,\ge\,2s'$. In both cases, the structure constants need to satisfy the algebra
\begin{equation}\eqlabel{4ptTspspsAlg}
 \sum_{a_P}\varepsilon_{b_1 b_4 a_P}\varepsilon_{a_P b_3 b_2}\:=\:
 \sum_{a_P}\varepsilon_{b_1 b_2 a_P}\varepsilon_{a_P b_3 b_4}.
\end{equation}
From the amplitude with two external states of spin-$s$ and two of spin-$s'$, the consistency condition becomes
\begin{equation}\eqlabel{4ptTspspsC2}
 \begin{split}
  &\left[1-(-1)^s\left(\frac{{\sf s}}{\sf t}\right)^s\right]\sum_{b_P}\varepsilon_{a_1 b_3 b_P}\varepsilon_{b_P b_4 a_2}+
   \frac{{\sf u}}{{\sf t}}\sum_{b_P}\varepsilon_{a_1 b_4 b_P}\varepsilon_{b_P b_3 a_2}\:=\\
  &\hspace{.5cm}=\:
   \left(\frac{{\sf s}}{{\sf t}}\right)^{s-1}\frac{\sf u}{\sf t}\sum_{a_P}\varepsilon_{a_1 a_2 a_P}\varepsilon_{a_P b_3 b_4}+
   \left(\frac{{\sf s}}{{\sf t}}\right)^{s-1}\frac{\sf u}{\sf t}\left(\frac{\kappa}{\kappa'}-1\right)
   \sum_{a_P}\varepsilon_{a_1 a_2 a_P}\varepsilon_{a_P b_3 b_4},
 \end{split}
\end{equation}
which is satisfied for $s\,=\,1$ and $\kappa'\,=\,\kappa$ with the algebra
\begin{equation}\eqlabel{4ptTspspsAlg2}
 \sum_{b_P}\epsilon_{a_1 b_4 b_P}\varepsilon_{b_P b_3 a_2}-\sum_{b_P}\varepsilon_{a_1 b_3 b_P}\varepsilon_{b_P b_4 a_2}\:=\:
 \sum_{a_P}\varepsilon_{a_1 a_2 a_P}\varepsilon_{a_P b_3 b_4},
\end{equation}
rediscovering the coupling of gluons with matter $s'\,\le\,1/2$.

Finally, we do not see any signature of a possible existence of higher-spin couplings with and without the introduction
of internal quantum numbers. In Section \eqref{subsec:s} we have seen
that, with our hypothesis, the self-interaction of particles with spin higher than two is trivial. This means that in
equation \eqref{4ptTspspsC2} the coupling constant $\kappa$ needs to be set to zero. It is easy to see that
for $s\,>\,2$ it is not possible to get a pure identity on the structure constants, without any function of the
kinematic variables.

For the sake of clarity, we summarise the consistent theories characterised by couplings with $s$-derivative interactions
in Table \ref{tab:SumS}.

\begin{table}[t]
 \centering
 \begin{tabular}{| l | l | l |}\hline
  $s$ & Conditions & Interactions\\\hline
  $s=0$ & $s'=\frac{1}{2}$, $\kappa=\kappa'$ & Yukawa\\\hline
  \multirow{3}{*}{$s=1$} &  $s'=0$, $\kappa=0$ & scalar QED and YM+scalars\\
  &  $s'=\frac{1}{2}$, $\kappa=0$ & QED and YM+fermions\\
  &  $s'=1$, $\kappa=\kappa'$ & YM\\\hline
  \multirow{3}{*}{$s=2$} &  $s'=0$, $\kappa=\kappa'$ & scalar GR\\
  &  $s'=\frac{1}{2}$, $\kappa=\kappa'$ & Fermion Gravity\\
  &  $s'=1$, $\kappa=\kappa'$ & Einstein-Maxwell\\
  &  $s'=\frac{3}{2}$, $\kappa=\kappa'$ & ${\cal N}=1$ supergravity\\
  &  $s'=2$, $\kappa=\kappa'$ & GR\\\hline
 \end{tabular}
 \caption{Summary of the theories characterised by couplings with $s$-derivative interactions.}
 \label{tab:SumS}
\end{table}

\subsection{Interactions with $3s$-derivatives}\label{subsec:3s}

In this section we discuss the second class of possible self-interacting theories. This class is characterised by
the following three-particle amplitudes
\begin{equation}\eqlabel{3ptAm3s}
 \begin{split}
  &M_{3}(1^{-s},2^{-s},3^{-s})\:=\:\kappa''\varepsilon_{a_1 a_2 a_3}
   \left(\langle1,2\rangle\langle2,3\rangle\langle3,1\rangle\right)^{s},\\
 &M_{3}(1^{+s},2^{+s},3^{+s})\:=\:\kappa''\varepsilon_{a_1 a_2 a_3}
   \left([1,2][2,3][3,1]\right)^s.
 \end{split}
\end{equation}
A simple dimensional analysis shows that such interactions would correspond, in the Lagrangian language, to three-point 
vertices with $3s$-derivative interactions. To our knowledge, this type of coupling typically emerges as an effective
interaction at low-energies ({\it e.g.} $F^3$, $R^3$). In the case of spin-$2$, a term of the type $R^3$ is the leading
counterterm at two-loops \cite{Goroff:1985th}, while in \cite{Wald:1986bj} it has been proposed an apparently classically
consistent theory which shows a $6$-derivative interaction, but which does not have general covariance.

A straightforward analysis of the helicity structure of the three-particle amplitudes \eqref{3ptAm3s} shows that
there is only one type of non-trivial four-particle amplitude: $M_4(1^{-s},2^{+s},3^{-s},4^{+s})$. Furthermore, from the
same analysis we can also infer that such an amplitude can have just one factorisation channel, the ${\sf u}$-channel.

This implies that a momentum-deformation involving the spinors of particle-$1$ and particle-$3$ generates
a one-parameter family of amplitudes which does not have any pole at finite location in the parameter $z$ and, 
as a consequence, the whole amplitude coincides with the contribution $\mathcal{C}_4^{\mbox{\tiny $(1,3)$}}$ from
the singularity at infinity, as it can be seen from eq. \eqref{AmplInt} by setting to zero the second term on the
right-hand-side.

Nevertheless, a momentum-deformation such as $(1,2)$ induces the presence of a pole at finite location, making the
generalised on-shell recursion relation \eqref{GenRR2} a meaningful mathematical representation again. However,
the amplitude does not have any factorisation channel other than the ${\sf u}$-channel, which invalidates our analysis of the collinear
limit used to fix the complex-UV parameter $\nu$ and the conditions on the zeroes.

If one thinks about the limits in which the S-matrix becomes trivial as a generic property of the scattering amplitudes
themselves, it is reasonable to assume that the condition \eqref{zeroes-4} on the zeroes can still hold
\begin{equation}\eqlabel{M43s0}
 P_{13}^2(z_0^{\mbox{\tiny $(l)$}})\:=\:-P_{1j}^2,
\end{equation}
where $j$ can be either $2$ or $4$, depending whether the $(1,2)$ or $(1,4)$-deformation is used. Applying these two
deformations plus the condition \eqref{M43s0}, we get
\begin{equation}\eqlabel{M4d3s}
 \begin{split}
  &M_4^{\mbox{\tiny $(1,2)$}}\:=\:(\kappa'')^2 \left(-1\right)^{\nu+1}\left(\frac{{\sf t}}{{\sf s}}\right)^{\nu+1}
   \frac{\langle1,3\rangle^{2s}[4,2]^{2s}}{{\sf u}}{\sf s}^s,\\
  &M_4^{\mbox{\tiny $(1,4)$}}\:=\:(\kappa'')^2 \left(-1\right)^{\nu+1}\left(\frac{{\sf s}}{{\sf t}}\right)^{\nu+1}
   \frac{\langle1,3\rangle^{2s}[4,2]^{2s}}{{\sf u}}{\sf t}^s,
 \end{split}
\end{equation}
where $\nu$ is, as usual, the complex-UV behaviour parameter, which is left generic here\footnote{One may think what 
happens if one assumes $\nu\,<\,0$ and, as a consequence, the ``weights'' are equal to one. The expression for the
amplitudes obtained through the two different deformations can be obtained from \eqref{M4d3s} by setting the term in round
brackets to one. The four particle test would imply $s\,=\,0$, which is not consistent with having just one factorisation channel.}. The four-particle test
\begin{equation}\eqlabel{4ptT3s}
 \frac{M_4^{\mbox{\tiny $(1,4)$}}}{M_4^{\mbox{\tiny $(1,2)$}}}\:=\:1\:=\:\left(\frac{\sf s}{\sf t}\right)^{2\nu+2-s}
\end{equation}
is satisfied if and only if $\nu\,=\,(s-2)/2$, with $s\,\ge\,2$. The four-particle amplitude can therefore be written
as 
\begin{equation}\eqlabel{M4d3sFin}
M_4(1^{-s},2^{+s},3^{-s},4^{+s})\:=\:(\kappa'')^2(-1)^{s/2}\frac{\langle1,3\rangle^{2s}[4,2]^{2s}}{{\sf u}}({\sf st})^{s/2}.
\end{equation}
A comment is now in order. The expression \eqref{M4d3sFin} for the four-point amplitude satisfies the only collinear limit
allowed (${\sf u}\,\rightarrow\,0$), as it is manifest from the on-shell construction, as well as it trivially satisfies
the soft limits, in which the amplitude vanishes. The form \eqref{M4d3sFin} seems to be consistent, at least for particles
with even spin. For particles with odd spin, the expression \eqref{M4d3sFin} shows branch points which in principle are not
expected.

Let us discuss in some detail the particular case of the spin-$2$ particle. From the expression \eqref{M4d3sFin}, 
the four-particle amplitude generated by the three-particle amplitudes \eqref{3ptAm3s} is given by
\begin{equation}\eqlabel{4ptAm3s}
 M_4(1^{-},2^{+},3^{-},4^{+})\:=\:-(\kappa'')^2\frac{\langle1,3\rangle^{4}[4,2]^{4}}{{\sf u}}{\sf st},
\end{equation}
and the contributions from the singularity at infinity for the deformations \eqref{BCFWs} and \eqref{BCFWs2} turn out to be
\begin{equation}\eqlabel{4ptC3s}
 \mathcal{C}_4^{\mbox{\tiny $(1,2)$}}\:=\:(\kappa'')^2\langle1,3\rangle^4 [4,2]^4 {\sf s},\qquad
 \mathcal{C}_4^{\mbox{\tiny $(1,4)$}}\:=\:(\kappa'')^2\langle1,3\rangle^4 [4,2]^4 {\sf t}.
\end{equation}
Such terms do not show any pole in momentum space. From a simple dimensional analysis, it is possible to infer that they 
correspond to four-particle $10$-derivative interactions. In principle, there is another way of looking at this type
of terms. By introducing a massive spin-$\tilde{s}$ particle, one can define further three-particle amplitudes involving
two spin-$2$ particles with different helicities and the auxiliary massive spin-$\tilde{s}$ particle. The introduction of 
this further coupling allows to define an effective four-particle amplitude with all the factorisation channels. 

Again, from dimensional analysis we can infer that this effective coupling $\tilde{\kappa}''$ should have mass-dimension 
$[\tilde{\kappa}'']\,=\,-4$, {\it i.e.} the auxiliary particle needs to be of spin-$5$. The terms \eqref{4ptC3s} should be
recovered then by taking both the mass of the auxiliary spin-$5$ particle and the effective coupling constant 
$\tilde{\kappa}''$ to be very large and keeping their ratio constant, which then needs to be identified as the 
three-particle coupling $\kappa''$.

Another choice can be to introduce the auxiliary massive particle by defining three-particle amplitudes involving two
spin-$2$ with the same helicities and the massive spin-$\tilde{s}$ particle. With the introduction of this effective
coupling, the four-particle amplitude still has just one factorisation channel to which two type of particles are
connected now. The new effective coupling constant has to have mass-dimension $0$, corresponding to a massive particle of
spin-$1$. As before, the original coupling constant is recovered as the ratio between the effective coupling constant and
the mass of the auxiliary particle in the limit where both of them are very large but their ratio is kept constant.

More generally, the structure of the contribution from the singularity at infinity is a polynomial of degree $s/2-1$ in 
the Mandelstam variables ${\sf s}$ and ${\sf t}$
\begin{equation}\eqlabel{4ptC3s2}
 \begin{split}
  &\mathcal{C}_4^{\mbox{\tiny $(1,2)$}}\,=\,\left(\kappa''\right)^2\langle1,3\rangle^{2s}[4,2]^{2s}\,{\sf s}^{s/2}\,
    {\sf P}_{\mbox{\tiny $(s/2-1)$}}^{\mbox{\tiny $(1,2)$}}({\sf s},{\sf t}),\\
  &\mathcal{C}_4^{\mbox{\tiny $(1,4)$}}\,=\,\left(\kappa''\right)^2\langle1,3\rangle^{2s}[4,2]^{2s}\,{\sf t}^{s/2}\,
    {\sf P}_{\mbox{\tiny $(s/2-1)$}}^{\mbox{\tiny $(1,4)$}}({\sf s},{\sf t}),
 \end{split}
\end{equation}
where the polynomials related to two deformations $(1,2)$ and $(1,4)$ are mapped into each other by the label exchange 
$2\,\longleftrightarrow\,4$. Also in these cases, a similar discussion to the one done in the spin-$2$ case holds. We
will further comment on these terms in Section \ref{sec:Loc}.

\subsection{Spin-$s$/spin-$s'$ interactions with $(2s'+s)$-derivatives}\label{subsec:2spps}

We consider now additional three-particle couplings with dimension $[\bar{\kappa}]\,=\,1-(2s'+s)$, which are characterised by
three-particle amplitudes with two particles of the same species and the same helicity:
\begin{equation}\eqlabel{3ptAm2sps}
 \begin{split}
  &M_{3}(1^{-s'},2^{-s'},3^{-s})\:=\:\bar{\kappa}\,\varepsilon_{b_1 b_2 a_3}
   \langle1,2\rangle^{2s'-s}\langle2,3\rangle^{s}\langle3,1\rangle^{s},\\
  &M_{3}(1^{+s'},2^{+s'},3^{+s})\:=\:\bar{\kappa}\,\varepsilon_{b_1 b_2 a_3}[1,2]^{2s'-s}[2,3]^{s}[3,1]^{s}.
 \end{split}
\end{equation}
Notice that the self-interaction coupling for spin-$s$ can have either dimension $s$ (Section \ref{subsec:s}) or
dimension $3s$ (Section \ref{subsec:3s}). Therefore, if we want to keep $s'$ generic rather than constrained to be
either $0$ or $s$, the coupling $\bar{\kappa}$ and the self-interaction one necessarily have different dimensions.
In the first instance, we can consider a theory defined just by the three-particle amplitudes in 
\eqref{3ptAm2sps}. In such a theory, it is possible to define just two non-trivial four-particle amplitudes, namely
$M_4(1^{-s'},2^{+s'},3^{-s'},4^{+s'})$ and $M_4(1^{-s},2^{+s},3^{-s'},4^{+s'})$. The first feature of such amplitudes
which is possible to infer from the helicity configurations \eqref{3ptAm2sps} is that they show just one factorisation
channel (with the label assignment chosen, it turns out to be the ${\sf u}$-channel), as it happens in the self-interacting
case with $3s$-derivatives. As a consequence, the same discussion holds about the action of the one-parameter deformations
on the amplitudes: Defining the one-parameter deformations shifting the spinors of particle-$1$ and particle-$3$,
there are no poles at finite location and the whole amplitude arises as a residue of the singularity at infinity;
while selecting other particles induces just one pole in the deformation parameter in the ${\sf u}$-channel (which makes 
manifest the only allowed factorisation channel), and consequently the analysis of the other collinear limit 
\eqref{zeroes-sum-4}  breaks down.

In what follows, we set to one the structure constants in \eqref{3ptAm2sps}. Strictly speaking, one can keep them trying to 
look for constraints on them. However, it is important to notice that, under all the useful momentum-deformations we can 
define, the same factorisation channel appears and, as a consequence, the four-particle amplitudes computed through the two 
different deformations would show the same factor involving the structure constants. Therefore, the four-particle test does 
not impose any constraint on the structure constants. We will do the same for the other cases which show just one
factorisation channel.

As for the class of theories discussed in Section \ref{subsec:3s}, we assume again that the zeroes are universally defined by
the condition \eqref{M43s0}. Choosing the deformations $(1,2)$ and $(1,4)$ on the amplitude 
$M_4(1^{-s'},2^{+s'},3^{-s'},4^{+s'})$, we get for $\nu\,\ge\,0$
\begin{equation}\eqlabel{4ptAm2sps}
 \begin{split}
  &M_4^{\mbox{\tiny $(1,2)$}}\:=\:\bar{\kappa}^2(-1)^{s+\nu+1}\left(\frac{\sf t}{\sf s}\right)^{\nu+1}
   \frac{\langle1,3\rangle^{2s'}[4,2]^{2s'}}{\sf u}{\sf s}^s,\\
  &M_4^{\mbox{\tiny $(1,4)$}}\:=\:\bar{\kappa}^2(-1)^{s+\nu+1}\left(\frac{\sf s}{\sf t}\right)^{\nu+1}
   \frac{\langle1,3\rangle^{2s'}[4,2]^{2s'}}{\sf u}{\sf t}^s.
 \end{split}
\end{equation}
The four-particle test leads to the condition \eqref{4ptT3s}, implying that $\nu\,=\,(s-2)/2$, with $s\,\ge\,2$. In the
case $\nu\,<\,0$, instead, the ``weights'' are equal to $1$, which amounts to set to one the functions of the kinematic
invariants which appear to the power of $\nu+1$ in eqs \eqref{4ptAm2sps}. It is straightforward to see that the 
four-particle test implies in this case that $s\,=\,0$.

In order to analyse the amplitude $M_4(1^{-s},2^{+s},3^{-s'},4^{+s'})$, again we can use the deformations $(1,2)$ and
$(1,4)$ to get
\begin{equation}\eqlabel{4ptAm2sps2}
 \begin{split}
  &M_4^{\mbox{\tiny $(1,2)$}}\:=\:
   \bar{\kappa}^2\frac{\langle1,3\rangle^s[3,2]^s\langle1,4\rangle^s[4,2]^s}{\sf u}
   \left(\frac{\langle1,3\rangle}{\langle1,4\rangle}\right)^{2s'}{\sf s}^{2s'-s}f_{13}^{\mbox{\tiny $(1,2)$}},\\
  &M_4^{\mbox{\tiny $(1,4)$}}\:=\:
   \bar{\kappa}^2\frac{\langle1,3\rangle^s[3,2]^s\langle1,4\rangle^s[4,2]^s}{\sf u}
   \left(\frac{\langle1,3\rangle}{\langle1,4\rangle}\right)^{2s'}{\sf t}^{2s'-s}f_{13}^{\mbox{\tiny $(1,4)$}},
 \end{split}
\end{equation}
where, similarly to \eqref{M4ssspsp12}, we are using $f_{13}^{\mbox{\tiny $(1,j)$}}$ to relate the ``weight'' of the 
${\sf u}$-channel to the correspondent deformation. Applying the four-particle test, the consistency condition reads
\begin{equation}\eqlabel{4ptT2sps}
 \frac{M_4^{\mbox{\tiny $(1,4)$}}}{M_4^{\mbox{\tiny $(1,2)$}}}\:=\:1\:=\:
  \frac{f_{13}^{\mbox{\tiny $(1,4)$}}}{f_{13}^{\mbox{\tiny $(1,2)$}}}\left(\frac{\sf t}{\sf s}\right)^{2s'-s}.
\end{equation}
For $\nu\,<\,0$, the ``weights'' are set to $1$ and the condition \eqref{4ptT2sps} is satisfied if and only if $s\,=\,2s'$.
Such a condition is clearly incompatible with the one obtained from the four-particle amplitude with all the external
states having spin-$s'$ in the case the latter vanishes at infinity, {\it i.e. $s\,=\,0$}: it would imply a 
self-interacting spin-$0$ theory with just one-channel. Instead, from the condition $s\,=\,2s'$ together with the constraint 
$s\,\ge\,2$ coming from the analysis of the four-particle amplitude with all the external states having spin-$s'$ in the 
case it does not vanish at infinity, one can easily deduce that this class of theories seem to be tree-level consistent if 
and only if $s'\,\ge\,1$.

For $\nu\,\ge\,0$, the condition on the zeroes \eqref{M43s0} implies that the ``weights'' in \eqref{4ptT2sps} are explicitly
\begin{equation}\eqlabel{f132spsR}
 \frac{f_{13}^{\mbox{\tiny $(1,4)$}}}{f_{13}^{\mbox{\tiny $(1,2)$}}}\:=\:\left(\frac{\sf s}{\sf t}\right)^{2(\nu+1)},
\end{equation}
and the four-particle test reads
\begin{equation}\eqlabel{4ptT2sps2}
 \frac{M_4^{\mbox{\tiny $(1,4)$}}}{M_4^{\mbox{\tiny $(1,2)$}}}\:=\:1\:=\:\left(\frac{\sf s}{\sf t}\right)^{2(\nu+1)-(2s'-s)},
\end{equation}
which is equal to $1$ if and only if $\nu\,=\,(2s'-s-2)/2\,\ge\,0$. Together with the condition $s\,\ge\,2$, the spin-$s$
is constrained to be in the range $s\,\in\,[2,2(s'-1)]$ and, as a consequence, $s'\,\ge\,2$. Saturating the lower bounds
$(s,s')\,=\,(2,2)$, one recovers the $3s$-derivative spin-$2$ self-interaction discussed in Section \ref{subsec:3s}.

\subsection{Spin-$s$/spin-$s'$ interactions with $|2s'-s|$-derivatives}\label{subsec:2spmns}

This class of theories is defined by three-particle amplitudes with the following two helicity configurations
\begin{equation}\eqlabel{3ptAm2spsM}
 M_{3}\left(1^{-s'},2^{-s'},3^{+s}\right),\qquad M_{3}\left(1^{+s'},2^{+s'},3^{-s}\right).
\end{equation}
As mentioned in Section \ref{subsec:GenConstr}, the massless three-particle amplitudes are zero on the real-sheet and this
requirement selects whether a certain three-particle amplitude is expressed just in terms of holomorphic/anti-holomorphic
spinors or it is a sum of both the holomorphic and anti-holomorphic term (this is valid for $0$-derivative interactions).

From \eqref{3ptam} it is easy to see that this condition changes the form of the three-particle amplitudes 
\eqref{3ptAm2spsM}, depending on whether $2s'-s$ is positive, negative or zero. We will analyse these three cases
separately.

First, a general comment on the structure of the four-particle amplitudes in this class of theories can already be made.
As it was the case for the $3s$-derivative self-interactions and the one discussed in the previous section, the helicity
configurations \eqref{3ptAm2spsM} of the three-particle amplitudes only allow to define non-trivial four-particle amplitudes
with only one factorisation channel, unless one of the particles involved in the scattering process is a scalar particle. 
Choosing the helicity assignments $M_{4}(1^{-s'},2^{+s'},3^{-s'}4^{+s'})$ and
$M_{4}(1^{-s},2^{+s},3^{-s'}4^{+s'})$, these two amplitudes factorise just in the ${\sf u}$-channel and ${\sf t}$-channel 
respectively. Therefore, also in this case we need to consider the condition \eqref{M43s0} to generically define the
zeroes of four-particle amplitudes.

\subsubsection{Case $s\,<\,2s'$}\label{subsubsec:sm2sp}

The two three-particle amplitudes defining the theories in this class are
\begin{equation}\eqlabel{3ptAm2spsMa}
 \begin{split}
  &M_{3}\left(1^{-s'},2^{-s'},3^{+s}\right)\:=\:
   \kappa\,\varepsilon_{b_1 b_2 a_3}\frac{\langle1,2\rangle^{2s'+s}}{\langle2,3\rangle^s\langle3,1\rangle^s},\\
  &M_{3}\left(1^{+s'},2^{+s'},3^{-s}\right)\:=\:\kappa\,\varepsilon_{b_1 b_2 a_3}\frac{[1,2]^{2s'+s}}{[2,3]^s[3,1]^s}.
 \end{split}
\end{equation}
Starting with the analysis of the amplitude $M_{4}(1^{-s'},2^{+s'},3^{-s'}4^{+s'})$, the on-shell representations generated
by the deformations $(1,2)$ and $(1,4)$ are given by
\begin{equation}\eqlabel{4ptAm2spsM}
 \begin{split}
  &M_{4}^{\mbox{\tiny $(1,2)$}}\:=\:\kappa^2(-1)^s\frac{f_{13}^{\mbox{\tiny $(1,2)$}}}{{\sf s}^s}
   \frac{\langle1,3\rangle^{2s'}[4,2]^{2s'}}{\sf u},\\
  &M_{4}^{\mbox{\tiny $(1,4)$}}\:=\:\kappa^2(-1)^s\frac{f_{13}^{\mbox{\tiny $(1,4)$}}}{{\sf t}^s}
   \frac{\langle1,3\rangle^{2s'}[4,2]^{2s'}}{\sf u},
 \end{split}
\end{equation}
where, as usual, the ``weights'' are one if the complex-UV parameter $\nu$ is negative, while if it is positive they
turn out to be
\begin{equation}\eqlabel{M4w2spsM}
 f_{13}^{\mbox{\tiny $(1,2)$}}\:=\:(-1)^{\nu+1}\left(\frac{\sf t}{\sf s}\right)^{\nu+1},\qquad
 f_{13}^{\mbox{\tiny $(1,4)$}}\:=\:(-1)^{\nu+1}\left(\frac{\sf s}{\sf t}\right)^{\nu+1}.
\end{equation}
The four-particle test imposes the following conditions 
\begin{equation}\eqlabel{4ptT2spsM}
 \begin{split}
  &\frac{M_{4}^{\mbox{\tiny $(1,4)$}}}{M_{4}^{\mbox{\tiny $(1,2)$}}}\:=\:1\:=\:\left(\frac{\sf s}{\sf t}\right)^s\,
   \qquad \nu\,<\,0,\\
 &\frac{M_{4}^{\mbox{\tiny $(1,4)$}}}{M_{4}^{\mbox{\tiny $(1,2)$}}}\:=\:1\:=\:
   \left(\frac{\sf s}{\sf t}\right)^{2(\nu+1)+s},\qquad \nu\,\ge\,0.
 \end{split}
\end{equation}
It is straightforward to see that, for $\nu\,<\,0$, the consistency condition is satisfied if and only if $s\,=\,0$, while
for $\nu\,\ge\,0$ the consistency requirement is never fulfilled. Let us assume then that $s\,=\,0$.

As far as the amplitude $M_{4}(1^{-s'},2^{+s'},3^{0},4^{0})$ is concerned, one easily sees that it is expected to 
factorise in the ${\sf u}$ and ${\sf t}$ channels. As a consequence, it is possible to define generalised on-shell
representations for which the analysis of the collinear limits holds, and fixes the conditions on the zeroes. With the
helicity assignments chosen above, such representations are generated by the
$(1,3)$ and $(1,4)$-deformations, where for particle-$1$ the anti-holomorphic spinor gets deformed while for particles
$3$ and $4$ the deformed spinor is the holomorphic one:
\begin{equation}\eqlabel{4ptAm2spsMb}
 \begin{aligned}
  &M_{4}^{\mbox{\tiny $(1,3)$}}\:=\:\kappa^2\,f_{14}\frac{\langle1,4\rangle^{2s'}[4,2]^{2s'}}{\sf t},
   &f_{14}\:=\:(-1)^{\nu+1}\left(\frac{\sf s}{\sf u}\right)^{\nu+1},\\
  &M_{4}^{\mbox{\tiny $(1,4)$}}\:=\:\kappa^2\,(-1)^{2s'}f_{13}\frac{\langle1,3\rangle^{2s'}[3,2]^{2s'}}{\sf u},
   &f_{13}\:=\:(-1)^{\nu+1}\left(\frac{\sf s}{\sf t}\right)^{\nu+1}.
 \end{aligned}
\end{equation}
The complex-UV parameter $\nu$ is fixed by the analysis of the collinear limits $[1,4]\,\rightarrow\,0$
and $[1,3]\,\rightarrow\,0$ for the representations respectively in the first and second line of eq \eqref{4ptAm2spsMb},
and, in both cases, it turns out to be given by $\nu\,=\,\delta+2h_1\,\equiv\,0$. One can immediately see how the
two expressions in \eqref{4ptAm2spsMb} do coincide when $\nu\,=\,0$. Thus, the amplitudes for such an amplitude takes the
following form
\begin{equation}\eqlabel{4ptAm2spsMf}
 M_4(1^{-s'},2^{+s'},3^{0},4^{0})\:=\:-\kappa^2\frac{\langle1,4\rangle^{2s'}[4,2]^{2s'}}{\sf t u}{\sf s}.
\end{equation}
This analysis seems to reveal that, at least at tree level, it is possible to define non-trivial couplings between
arbitrary spin-$s'$ particles and a scalar. 

\subsubsection{Case $s\,>\,2s'$}\label{subsubsec:sM2sp}

For $s\,>\,2s'$, the relevant three-particle amplitudes acquire the following form
\begin{equation}\eqlabel{3ptAm2spsP}
 \begin{split}
  &M_3\left(1^{-s'},2^{-s'},3^{+s}\right)\:=\:\kappa\,\varepsilon_{b_1 b_2 a_3}\frac{[2,3]^{s}[3,1]^{s}}{[1,2]^{2s'+s}},\\
  &M_3\left(1^{+s'},2^{+s'},3^{-s}\right)\:=\:\kappa\,\varepsilon_{b_1 b_2 a_3}
   \frac{\langle2,3\rangle^{s}\langle3,1\rangle^{s}}{\langle1,2\rangle^{2s'+s}}.
 \end{split}
\end{equation}
Focusing, at first, on the amplitude with four external states with spin $s'$ ($s'\,\neq\,0$; the case $s'=0$ is not 
consistent when the four-particle amplitudes show just one factorisation channel, as we saw in the previous sections), we 
generate its on-shell representations through the following deformations
\begin{equation}\eqlabel{BCFW2spsP}
 \lambda^{\mbox{\tiny $(1)$}}(z)\:=\:\lambda^{\mbox{\tiny $(1)$}}+z\lambda^{\mbox{\tiny $(j)$}},\qquad
 \tilde{\lambda}^{\mbox{\tiny $(j)$}}(z)\:=\:\tilde{\lambda}^{\mbox{\tiny $(j)$}}-z\tilde{\lambda^{\mbox{\tiny $(1)$}}},
 \quad j\,=\,2,4.
\end{equation}
Such representations are explicitly
\begin{equation}\eqlabel{4ptAm2spsP}
 \begin{split}
  &M_{4}^{\mbox{\tiny $(1,2)$}}\:=\:\kappa^2 (-1)^s f_{13}^{\mbox{\tiny $(1,2)$}}
   \frac{\langle1,3\rangle^{2s'}[4,2]^{2s'}}{\sf u}\frac{{\sf s}^s}{{\sf u}^{4s'}},\\
  &M_{4}^{\mbox{\tiny $(1,4)$}}\:=\:\kappa^2 (-1)^s f_{13}^{\mbox{\tiny $(1,4)$}}
   \frac{\langle1,3\rangle^{2s'}[4,2]^{2s'}}{\sf u}\frac{{\sf t}^s}{{\sf u}^{4s'}}.
 \end{split}
\end{equation}
Imposing the four-particle test for $\nu\,<\,0$ implies, as in the previous section, that the particle of spin $s$ has
to be a scalar. However, we are focusing on theories where the spins of the two particles involved are constrained
by the inequality $s\,>\,2s'$, with $s'\,\neq\,0$. The case in which $s\,=\,0$ is therefore ruled out. For $\nu\,\ge\,0$,
the consistency condition becomes
\begin{equation}\eqlabel{4ptT2spsP}
 \frac{M_{4}^{\mbox{\tiny $(1,4)$}}}{M_{4}^{\mbox{\tiny $(1,2)$}}}\:=\:1\:=\:\left(\frac{\sf s}{\sf t}\right)^{2(\nu+1)-s},
\end{equation}
which is satisfied for $\nu\,=\,s/2-1$. One comment is now in order. The consistency condition \eqref{4ptT2spsP}
does not impose any constraint on the spin $s'$ of the external particles. However, it can already be noticed from
the expressions in \eqref{4ptAm2spsP} that for $s'\,\neq\,0$, the ${\sf u}$-channel shows a multiple pole. This
 breaks one of the fundamental hypothesis on which our construction is based, so necessarily $s'\,=\,0$. 
Strictly speaking, the amplitude with four-external spin-$0$ states shows two BCFW-channels under any of the two
deformations considered. Anyhow, it is also worth to notice that this case would reduce to the one discussed in
Section \ref{subsec:s} with the external states being all scalars. 

\subsubsection{Case $s\,=\,2s'$}\label{subsubsec:su2sp}

Finally, the case $s\,=\,2s'$ corresponds to a class of theories with $0$-derivative interactions. The three-particle
amplitudes are characterised by being a linear combination of the holomorphic and anti-holomorphic terms:
\begin{equation}\eqlabel{3ptAm2sp2sp}
 \begin{split}
  &M_3(1^{-s'},2^{-s'},3^{+2s'})\:=\:
   \kappa^{\mbox{\tiny H}}\varepsilon_{b_1 b_2 a_3}
    \frac{\langle1,2\rangle^{4s'}}{\langle2,3\rangle^{2s'}\langle3,1\rangle^{2s'}}+
   \kappa^{\mbox{\tiny A}}\varepsilon_{b_1 b_2 a_3}\frac{[2,3]^{2s'}[3,1]^{2s'}}{[1,2]^{4s'}},\\
  &M_3(1^{+s'},2^{+s'},3^{-2s'})\:=\:
   \kappa^{\mbox{\tiny H}}\varepsilon_{b_1 b_2 a_3}
    \frac{\langle2,3\rangle^{2s'}\langle3,1\rangle^{2s'}}{\langle1,2\rangle^{4s'}}+
   \kappa^{\mbox{\tiny A}}\varepsilon_{b_1 b_2 a_3}\frac{[1,2]^{4s'}}{[2,3]^{2s'}[3,1]^{2s'}}.
 \end{split}
\end{equation}
As usual, we would need to analyse two four-particle amplitudes, one having all external states with spin $s'$ and
the other one having two external states with spin $s'$ and two with spin $2s'$. Let us start with the four-particle
amplitude with all the external states with spin $s'$, for which we choose the helicity assignment 
$M_4(1^{-s'},2^{+s'},3^{-s'},4^{+s'})$. The on-shell representations induced by the deformations
\begin{equation}\eqlabel{BCFW2sp2sp}
 \tilde{\lambda}^{\mbox{\tiny $(1)$}}(z)\:=\:\tilde{\lambda}^{\mbox{\tiny $(1)$}}-z\tilde{\lambda}^{\mbox{\tiny $(j)$}},
  \quad
 \lambda^{\mbox{\tiny $(j)$}}(z)\:=\:\lambda^{\mbox{\tiny $(j)$}}+z\lambda^{\mbox{\tiny $(1)$}},\qquad
  j\,=\,2,4
\end{equation}
take the form
\begin{equation}\eqlabel{4ptAm2sp2sp}
 \begin{split}
  &M_{4}^{\mbox{\tiny $(1,2)$}}\:=\:\hat{\kappa}^2(-1)^{2s'}\frac{f_{13}^{\mbox{\tiny $(1,2)$}}}{\sf u}
    \frac{\langle1,3\rangle^{2s'}[4,2]^{2s'}}{{\sf s}^{2s'}},\\
  &M_{4}^{\mbox{\tiny $(1,4)$}}\:=\:\hat{\kappa}^2(-1)^{2s'}\frac{f_{13}^{\mbox{\tiny $(1,4)$}}}{\sf u}
    \frac{\langle1,3\rangle^{2s'}[4,2]^{2s'}}{{\sf t}^{2s'}},
 \end{split}
\end{equation}
where $\hat{\kappa}\,\overset{\mbox{\tiny def}}{=}\,\kappa^{\mbox{\tiny H}}+\kappa^{\mbox{\tiny A}}$ is the effective
three-particle coupling. From \eqref{4ptAm2sp2sp}, the four-particle test leads to the constraint
\begin{equation}\eqlabel{4ptT2sp2sp}
 \frac{M_{4}^{\mbox{\tiny $(1,4)$}}}{M_{4}^{\mbox{\tiny $(1,2)$}}}\:=\:1\:=\:
  \frac{f_{13}^{\mbox{\tiny $(1,4)$}}}{f_{13}^{\mbox{\tiny $(1,2)$}}}\left(\frac{\sf s}{\sf t}\right)^{2s'}.
\end{equation}
For $\nu\,<\,0$ the ``weights'' are $1$, while for $\nu\,\ge\,0$ their ratio in \eqref{4ptT2sp2sp} turns out to be
$({\sf s}/{\sf t})^{\nu+1}$. It is straightforward to conclude that the constraint \eqref{4ptT2sp2sp} can be never satisfied for 
$s'\,\neq\,0$.

\section{On-Shell Representation, High-Spin Theories and Locality}\label{sec:Loc}

Let us now discuss the results of the previous sections, focusing on the possibility of describing theories involving 
higher-spin particles. The consistency conditions on the four-particle amplitudes seem to provide us with some classes of 
theories where the particles have their spin partially constrained. Let us go through all the cases systematically:
\begin{enumerate}
 \item Self-interactions with $s$-derivatives (Section {\ref{subsec:s}}): These are theories characterised by three-particle
       amplitudes with helicity configurations $(\mp s,\mp s,\pm s)$, which is the class where  
       $\lambda\phi^3$, Yang-Mills and GR fall in. The consistency requirement \eqref{4ptTest} rules out the possibility
       of having any self-interacting theory -- with or without internal quantum numbers -- with spin $s\,>\,2$;
 \item Spin-$s$/Spin-$s'$ interactions with $s$-derivatives (Section {\ref{subsec:ssp}}): The three-particle amplitudes
       characterising this class of theories have helicity configurations $(\mp s',\pm s', \mp s)$, with the spin-$s$
       particle playing the role of mediator. The four-particle test on the generalised on-shell recursion relation allows 
       us to rediscover the couplings of gravitons and photons/gluons with particles with lower spin, with their
       algebra structure if any, as well as the Yukawa interactions. Again, with our assumptions, we do not find in this 
       class of theories any signature of the possible existence of a scattering process involving high-spin particles;
 \item Self-interactions with $3s$-derivatives (Section {\ref{subsec:3s}}): These are theories defined through 
       three-particle amplitudes with helicity configurations $(\mp s, \mp s, \mp s)$. They have the peculiar feature
       of allowing just one factorisation channel in four-particle amplitudes. The consistency condition seems not to 
       forbid self-interaction of particles with even spin higher or equal to two;
 \item Spin-$s$/Spin-$s'$ interactions with $(2s'+s)$-derivatives (Section \ref{subsec:2spps}): The three-particle 
       amplitudes defining the theories in this class have helicity configurations $(\mp s', \mp s', \mp s)$. The
       non-trivial couplings that seem to appear are interactions between particles with spin $s'$ and particles with spin $s\,=\,2s'$,
       with $s'\,\ge\,1$, and spin-$s$/spin-$s'$ interactions with $s\,\in\,[2,\,2(s'-1)]$;
 \item Spin-$s$/Spin-$s'$ interactions with $|2s'-s|$-derivatives (Section \ref{subsec:2spmns}): In this last case,
       the fundamental building-blocks are three-particle amplitudes with helicity structure $(\mp s', \mp s', \pm s)$.
       The consistency analysis seems to reveal the possibility of non-trivial couplings between a scalar ($s\,=\,0$) and
       particles of spin $s'\,>\,0$.
\end{enumerate}
Let us comment in some detail on the last three classes of theories. All of them have in common the feature of possessing
just one factorisation channel for four-particle amplitudes. 

As we commented previously, this means that there are two 
classes of momentum-deformation that in principle we can define in order to construct the on-shell representation of
such theories. 

One does not induce any pole in the parameter $z$ at finite location: the amplitude coincides with the
contribution from the singularity at infinity. In this case our generalised construction, which is based on
the knowledge of a subset of poles and zeroes, breaks down. 

The second class of deformations instead induces a pole
for the amplitude in the parameter $z$, corresponding to the same channel in which the amplitude factorises in
momentum space. Therefore, it is possible to formally write down the on-shell representation \eqref{GenRR2} which
shows explicitly the only allowed factorisation channel. The catch here is that the zeroes cannot be fixed by
the collinear limit analysis suggested in \cite{Benincasa:2011kn}. 

We overcame this issue by generalising the condition \eqref{zeroes-4} on the zeroes also to these cases, thinking about
the zeroes of an amplitude on the same footing as the poles, {\it i.e.} as ``universal'' limits in momentum-space. 
Obviously, this is a strong assumption which needs to be checked. The expressions we got using this assumption do not
seem to show any obvious inconsistency (unless somehow causality breaks down). 

Furthermore, especially in these cases, it is not obvious that the four-particle consistency at tree level implies 
consistency of the tree-level scattering amplitude for any number of external states. So, a possible way to check if there is some 
pathology arising would be to apply the generalisation of the test \eqref{4ptTest} at $5$-particle level and try to argue if the consistency 
conditions may break down. One could also check whether combining different three-particle couplings it could possible to 
define a theory which is tree-level consistent\footnote{Obviously, the tree-level consistency would not be the end of the
story. A theory could show pathologies at loop level, {\it e.g.} ghosts. However, if a theory built out of a single
type of coupling shows ghosts, there might be the possibility to cancel them by considering several types of such 
couplings.}. 

An interesting feature of these classes of theories is the form of the contribution from the singularity
at infinity $\mathcal{C}_{4}^{\mbox{\tiny $(i,j)$}}$. Let us think of the amplitude as the sum of a standard BCFW-term
and $\mathcal{C}_{4}^{\mbox{\tiny $(i,j)$}}$, rather than in the form \eqref{GenRR2}. As it can be seen from the expressions
\eqref{4ptC3s2} for the $3s$-derivative interaction (it can be explicitly checked also for the other cases), such terms
do not show any pole in the Mandelstam variables. Certainly, this is not unexpected given that there is no other 
factorisation channel than the one contained in the standard BCFW-term. 

As a consequence $\mathcal{C}_{4}^{\mbox{\tiny $(i,j)$}}$ has the structure of a contact interaction. Dimensional analysis
on \eqref{4ptC3s2} reveals that such a contact interaction has $2(3s-1)$-derivatives. It is very likely that, provided
that the theory is tree-level consistent, the $n$-particle amplitude would show an $n$-particle contact term in this
on-shell representation. Focusing, again just as an example, on the $3s$-derivative interaction case, the number of 
derivatives of such a contact interaction turns out to be $L_n\,=\,(3s-2)n-6(s-1)$. 

More generally, given a theory with $\delta$-derivative three-particle couplings, the $n$-particle contact term turns out 
to be characterised by $L_n$ derivatives, with $L_n\,=\,(\delta-2)n-2(\delta-3)$. In the cases of relevance, it is easy to 
see that $L_n$ increases with $n$. Therefore, these theories seem to be endowed with higher-derivative interaction terms 
with the increase of the number of external states. This can be interpreted as a signature of non-locality for these theories.

The theories we have discussed so far passed the four-particle test without having internal quantum numbers. 
It is possible to impose the four-particle test keeping the structure
constants defined for the three-particle amplitudes. The catch is that both on-shell representations generated through 
two different momentum deformations show the same channel (it cannot be otherwise given that these theories have just
one factorisation channel allowed) and, therefore, the same factor containing the structure constants appear in both
sides of the condition \eqref{4ptTest}. Thus it does not seem possible to obtain any constraint for the structure 
constants algebra, which would imply the existence of a non-abelian internal symmetry. One could think of endowing the
contact interaction with some internal structure, but this seems to be quite unnatural.

In the known example of supposedly consistent high-spin couplings in flat space, non-locality, a non-abelian structure and
the existence of a large number of propagating degrees of freedom ({\it e.g.} \cite{Taronna:2011kt} and references therein)
seem to be key features to be able to define high-spin interactions.

Very likely, the theories involving spin higher than two we have discovered are not going to be consistent (we took
just the simplest possible interactions with very few particles involved). However, the main aim of this discussion was to
check if, in some sense, the generalised on-shell representation could be a good arena for consistently look for high-spin
theories\footnote{An attempt to use the BCFW construction for high-spin theories was done in \cite{Fotopoulos:2010ay}.}. 
Surprisingly enough, starting from four basic hypothesis (Poincar{\'e} invariance, analyticity of the S-Matrix,
locality, existence of $1$-particle states) we arrived to the possibility of seeing the breaking down of the locality
requirement through the contact terms generated as the contribution from the singularity at infinity.

\section{Conclusion}\label{sec:Concl}

Using the generalised on-shell representation \eqref{GenRR2} and the tree-level consistency condition proposed in
\cite{Benincasa:2007xk}, we explored the space of tree-level consistent interacting theories of massless particles. 

We found it useful to classify the interactions through the dimensionality of the three-particle coupling constants. 
Each class defined in this way then turns out to be characterised by very specific helicity configurations of the 
three-particle amplitudes, which induce the factorisation properties in the amplitudes with higher number of external
states. 

An essential key is the extension of the notion of constructibility implied by the generalised on-shell recursion relations
\eqref{GenRR2}. They allow to express the tree-level scattering amplitudes of any theory as a product of scattering 
amplitudes with a lower number of external states, weighted by a factor dependent on a subset of zeroes 
\cite{Benincasa:2011kn}. These ``weights'' can be fixed by requiring that the representation \eqref{GenRR2} factorises 
properly under the relevant collinear/multi-particle limits \cite{Benincasa:2011kn}. The crucial point in this analysis
is related to the collinear limit involving the two particles whose momenta have been deformed. Indeed, such a channel does
not explicitly appear in the recursion relation. This limit is instead translated into a soft limit of (at least) one of
the deformed particles and, therefore, the collinear singularity appears as a soft singularity in the case of the standard 
BCFW recursion relation \cite{Schuster:2008nh, Benincasa:2011kn}. 

This points towards a strong link between the soft limits of the three-particle amplitudes and the complex-UV behaviour.
First, in the collinear-limit analysis we just referred to, it turns out that the only on-shell diagrams contributing are
those containing three-particle amplitudes with one of the two deformed momenta, depending whether the holomorphic or 
anti-holomorphic limit is taken. Furthermore, taking the collinear limit induces the deformed momentum appearing in the
three-particle amplitudes to become soft. 

Now, we know that the standard BCFW representation is allowed if and only if the one-parameter family of
amplitudes induced by the momentum deformation vanishes as the parameter is taken to infinity and fails if the amplitude
behaves at infinity as a constant or diverges. 

In the same fashion, such a representation is allowed if the deformed particle becoming soft realises the appropriate
singularity through the three-particle amplitudes it belongs to. When this soft limit is not instead able to induce the
correct pole, the standard BCFW representation fails. 

Thus, the soft-behaviour of the three-particle amplitude can be used as a criterion to establish the validity of the 
standard BCFW construction (the usual constructibility notion) or, more generically, to establish whether the ``weights''
in the generalised on-shell construction are $1$ or they are expressed in terms of a subset of the zeroes. 

We analysed in detail the soft-behaviour of the three-particle amplitudes, obtaining the exponent as a function of the
number of derivatives of the relevant three-particle interaction and of the helicity of the soft particle. Amazingly,
this exponent agrees with the large-$z$ behaviour one in all the known cases. 

Hence, the analysis of the soft-behaviour of the three-particle amplitude provides with a simple way to understand
if the amplitudes with a higher number of external states are constructible in the usual sense or in the generalised one.

Interestingly, the generalised on-shell representation and the consistency test allows us to detect all the known couplings,
their gauge algebra, if any, and the relation between different couplings with the same mass dimension. As already noticed
in \cite{Benincasa:2007xk} for Yang-Mills and $\mathcal{N}\,=\,1$ supergravity, this implies that structures like the
gauge algebra and supersymmetry do not necessarily need to be imposed a priori but they arise from a consistency condition.

All the known theories turn out to belong to the classes of couplings defined by the three-particle amplitudes with
helicity structure $(\mp s, \mp s, \pm s)$ and $(\mp s', \pm s', \mp s)$, which are all interactions with $s$-derivative,
where $s$ is the spin of the mediator.

However, those are not the only couplings allowed at three-particle level. The other couplings have the peculiar feature
of inducing just a single factorisation channel in the four-particle amplitudes. Even if the generalised on-shell recursive
relation remains valid under some deformations, the collinear limit analysis which fixes the zeroes of the amplitude
(and therefore the ``weights'') breaks down. Thinking about the zeroes as a result of some ``universal kinematic limit'',
as it happens for the poles, we assume that the condition on the zeroes previously found can be extended to these cases
as well. As emphasised in the text, this is a strong heuristic assumption which needs to be checked.

The results of the consistency test on these classes of theories do not seem to show any obvious pathology, but indeed they 
would allow for some interesting results, {\it e.g.} $6$-derivative self-interactions of spin-$2$ particles, high-derivative
interactions among high-spin particles and between high-spins and a scalar. Interestingly, all these theories show
four-particle amplitudes whose term coming from the singularity at infinity has the structure of a contact interaction with
a number of derivatives higher than the ones of the related three-particle interaction. Moreover, a simple dimensional
analysis show that the appearance of such a term at $n$-particle level would have $L_n\,=\,(\delta-2)n-2(\delta-3)$ 
derivatives, with $\delta$ being the number of derivatives of the three-particle interaction. For the cases of interest,
$L_n$ increases with the number of external states $n$. This a signature of non-locality.

Now, it is obvious that our analysis does not allow us to claim that we were able to detect consistent interacting theories
of high spin in flat-space. Taking for good our assumption about the zeroes of the four-particle amplitudes with just
one factorisation channel, the four-particle consistency does not necessarily implies consistency of the scattering 
amplitude for any number of external states. In addition, the apparently consistent tree-level amplitudes we found
seem to be characterised by the breaking of one of our basic assumption, locality.

However, this is at the same time a striking result. It is an indication that in order to look for potentially consistent
theories of high spin, locality needs to be dropped as a fundamental requirement. From the S-matrix perspective, this
 would lead to dramatic effects, because it would change the analytic structure of the amplitudes. For example, it is
not clear if it would be still consistent to consider the tree level as characterised by poles only.

It would be interesting to understand the consequences of weakening this requirement on general grounds. In a more 
short-term view, it would be great to test the amplitudes we found by developing a rigorous way to fix the
``weights'' of the generalised on-shell representation for such theories. If this proves our assumption to be correct,
it would be interesting to perform a consistency check on amplitudes with a higher number of external states. Finally,
another direction would be to make a systematic study of these couplings were more degrees of freedom are propagating.

\section*{Acknowledgement}

P.B. would like to thank the organisers of the workshops ``String Theory'' and ``Gravity - New Perspective from strings and
higher dimensions'' held at the Center de Ciencias ``Pedro Pascual'' in Benasque, where the very last stages of this work
have been carried out. This work is funded in part by MICINN under grant FPA2008-01838, by the Spanish Consolider-Ingenio 
2010 Programme CPAN (CSD2007-00042) and by Xunta de Galicia (Conseller{\'i}a de Educac{\'i}on, grant INCITE09 206 121 PR and
grant PGIDIT10PXIB206075PR) and by FEDER. P.B. is supported as well by the MInisterio de Ciencia e INNovaci{\'on} through 
the Juan de la Cierva program. E.C. is supported by a Spanish FPU fellowship, and would like to thank Perimeter Institute for hospitality and the FRont Of Galician-speaking 
Scientists for unconditional support.

\bibliographystyle{utphys}
\bibliography{amplitudesrefs}	

\end{document}